\documentclass[11pt]{article}
\usepackage{tikz}
\usepackage{subcaption}
\usepackage{eurosym}
\usepackage{amsthm}
\newtheorem{proposition}{Proposition}
\usepackage{amsmath}
\usepackage{amssymb}
\usepackage{mathtools}
\usepackage{graphicx}
\usepackage{setspace}
\usepackage{sectsty}
\usepackage{soul}
\usepackage[authoryear]{natbib}
\usepackage[margin=1in]{geometry}
\usepackage{color}
\usepackage{bbm}
\usepackage{booktabs}
\usepackage{pdflscape}
\usepackage{dcolumn}
\usepackage{threeparttable}
\usepackage{siunitx}
\usepackage{array}
\usepackage{appendix}
\usepackage{enumitem}
\usepackage{placeins}
\usepackage{fancyhdr}
\usepackage[breaklinks=true]{hyperref}
\usepackage{cases}
\usepackage{etoolbox}
\usepackage{titlesec}
\usepackage{ragged2e}
\usepackage{lmodern,textcomp}
\usepackage[utf8x]{inputenc}
\usepackage{longtable} 
\usepackage{import}
\usepackage{minitoc}

\theoremstyle{plain}                     
  \newtheorem{theorem}{Theorem}          
  \newtheorem{lemma}[theorem]{Lemma}

\theoremstyle{definition}                

\theoremstyle{remark}                    
  
\begin{document}
\doparttoc 

    \begin{titlepage}
\renewcommand{\baselinestretch}{1}
	\title{
		\sc{
			\Large{
            Monetary Policy, Uncertainty, and Credit Supply
			}%
		}%
	\thanks{We would like to thank Olivier de Bandt, Jean Barthélemy, Thorsten Beck, Agnès Benassy-Quéré, Antoine Camous, Laurent Clerc, Hans Degryse, Jose Fellmann, David Gauthier, Adam Golinski, Christoph Grosse-Steffen, Sophie Guilloux-Nefussi, David Kremer, Stéphane Lhuissier, Franziska Maruhn, Magali Marx, Nicola Pavanini, Adrian Penalver, Paul Reimers, Arthur Saint-Guilhem, Manuel Schick, Jean-David Sigaux, and Elias Wolf for their comments. The views expressed in this paper are solely those of the authors and do not necessarily reflect the views of their past, present, or future employers.}
	}			
\author{
\textbf{Eric Vansteenberghe}%
\thanks{
Banque de France: \href{mailto:eric.vansteenberghe@banque-france.fr}{\nolinkurl{eric.vansteenberghe@banque-france.fr}}
}
}
	
	\date{}
	
	\maketitle

\abstract{\noindent
This paper investigates how dispersion in banks’ subjective inflation forecasts is a channel of the transmission of monetary policy to credit supply. We extend the Monti--Klein model of monopolistic banking by incorporating risk aversion, subjective beliefs, and ambiguity aversion. The model predicts that greater inflation uncertainty or asymmetry in beliefs raises equilibrium loan rates and amplifies credit rationing. Using AnaCredit loan-level data for France, we estimate finite-mixture density regressions that allow for latent heterogeneity in loan pricing. Empirically, we find that higher subjective uncertainty and asymmetry both increase average lending rates and skew their distribution, disproportionately affecting financially constrained firms in the right tail. Quantitatively, moving from the 25th to the 75th percentile of our indicators raises average borrowing costs by more than 10 basis points, which translates into roughly 0.5 billion euros of additional annual interest expenses for non-financial corporations. By contrast, forecast disagreement has a weaker and less systematic effect. Taken together, these results show that uncertainty and asymmetry in inflation expectations are independent and powerful drivers of credit conditions, underscoring their importance for understanding monetary policy transmission through the banking sector.
\label{abstract}
}

    
\thispagestyle{empty}
\end{titlepage}

    \section{Introduction}

This paper examines how banks' subjective inflation forecasts shape their response to monetary policy and ultimately influence credit supply. Monetary policy decisions and communications can increase forecast dispersion, understood as the combination of uncertainty, disagreement, and asymmetry in expectations. Higher dispersion prompts banks to adjust lending terms in a precautionary manner, typically through higher interest rates and tighter credit supply. These responses generate significant costs for firms and, more broadly, weaken the transmission of monetary policy through the banking sector. By incorporating banks’ risk aversion and belief formation into models of credit provision, we show that greater forecast dispersion raises corporate borrowing costs and restricts access to credit. The macroeconomic significance of these mechanisms highlights the need for central banks to consider how communication strategies and policy design can reduce forecast dispersion and preserve the effectiveness of monetary policy transmission.

A credible central bank commitment to an inflation target serves to anchor expectations, thereby reducing forecast dispersion and stabilizing economic outcomes \citep{kydland1977rules,barro1983rules}. The inflation forecasts of financial intermediaries, particularly at the medium-term horizon, provide a valuable gauge of how firmly expectations remain anchored \citep{woodford2005interest}. Systematic policy responses to deviations of expected inflation from the target are commonly formalized in reaction functions such as the \citet{taylor1993discretion} rule. Beyond the adjustment of policy instruments, central banks may also rely on communication and information disclosure to influence the dispersion of inflation expectations and thereby reinforce the effectiveness of monetary policy transmission \citep{clarida1999science}. While these mechanisms highlight the central role of expectations in monetary policy design, they leave open the question of how forecast dispersion affects banks’ own credit supply decisions. Addressing this gap motivates our theoretical framework.

Our theoretical framework extends the classic Monti--Klein model of monopolistic banking by integrating insights from the credit channel \citep{bernanke1995inside,stiglitz1981credit}, the risk-taking channel \citep{kashyap2023monetary}, and the deposits channel \citep{drechsler2017deposits}. Departing from the standard assumptions of rational expectations and risk neutrality, we explicitly incorporate banks' risk aversion and subjective belief formation. This extension allows us to examine how deviations from rational expectations, and in particular the dispersion of inflation beliefs, affect optimal lending rates and credit supply. The model predicts that greater dispersion of inflation beliefs raises equilibrium lending rates through its impact on banks' expected profits, which constitutes our main testable hypothesis. Extending the framework to include borrower risk, as in \citet{stiglitz1981credit}, we show that higher inflation uncertainty amplifies credit rationing. Moreover, when banks are ambiguity-averse \citep{klibanoff2005smooth}, both effects not only persist but are reinforced, highlighting the central role of belief dispersion in the transmission of monetary policy.

Building on these theoretical predictions, our empirical framework is designed to test how dispersion in inflation beliefs affects loan pricing. We control for observed loan characteristics (maturity, volume, probability of default, lender traits, and borrower sector) as well as the broader macroeconomic environment, including the monetary policy stance. To estimate these effects, we employ a finite mixture of generalized linear models, which accommodates latent heterogeneity in the data and facilitates the identification of distinct pricing strategies or borrower groups as in \citet{lacroix2008assessing}. Our main results indicate that dispersion in inflation beliefs significantly distorts the distribution of interest rates offered to non-financial corporations. When subjective dispersion increases, the annualized interest rates charged to firms rise not only in their median but also display greater positive variance. This distortion is particularly consequential for monetary policy, as small firms, which are more likely to be financially constrained \citep{beaumont2025collateral}, tend to be disproportionately represented in the right tail of the interest rate distribution.

To further assess the robustness of our findings, we restrict the sample to \emph{newly issued overdraft facilities}. These credit lines constitute a particularly informative subset from an econometric standpoint. By construction, overdrafts are short-term, often renewed at high frequency, and typically uncollateralized. This design largely eliminates the joint determination of maturity, collateral, and loan volume that characterizes other credit instruments. In this restricted setting, the main source of variation in the contractual interest rate lies in banks’ own pricing decisions rather than in borrower demand or contract structure. This feature allows us to isolate the component of lending rates that reflects the bank’s cost assessment and risk pricing under uncertainty. Specifically, we examine the \emph{cost spread}—defined as the difference between the nominal interest rate charged on new overdrafts and the corresponding risk-free short-term benchmark. When controlling for borrower risk (probability of default), borrower industry, location, and size; macroeconomic conditions (GDP growth and policy stance); and bank fixed effects, the residual variation in this cost spread captures the bank’s discretionary pricing margin. Regressing this margin on the indicators of forecast dispersion (NIU, ASI, and disagreement) thus provides a sharper test of whether inflation uncertainty affects banks’ own rate-setting behavior.

\paragraph{Related literature} 
Recent research has examined whether survey-based inflation forecasts remain anchored to central bank targets, particularly in the aftermath of pandemic-induced inflation surges. \citet{coibion2022monetary} find that direct communication lowers long-term inflation expectations and affects spending. Communication also influences the uncertainty embedded in subjective probability distributions: announcements shape financial markets not only through the level of policy rates, but through their impact on agents' beliefs and uncertainty about the future path of policy and the economy \citep{beber2006effect,bauer2022market,ehrmann2019can}. Extending this literature, \citet{vansteenberghe2025uncertain} introduce two distribution-based indicators, the Normalized Uncertainty (NU) and the Asymmetry Strength Index (ASI), and show that monetary policy transmission is state-dependent. Forward-guidance and quantitative-easing shocks re-anchor expectations only when uncertainty is low, conventional rate shocks destabilize them, and higher normalized uncertainty systematically depresses growth forecasts.

Research has also emphasized how monetary policy transmits through the banking sector. \citep{bernanke1995inside} highlight the \textit{credit channel}, emphasizing monetary policy transmission through borrower balance sheets and banks' lending capacity. Tighter policy raises external finance premiums and reduces collateral values, constraining credit demand, while funding pressures limit banks' loan supply, particularly affecting financially weaker institutions. Under asymmetric information, banks ration credit instead of raising interest rates excessively, due to adverse selection and moral hazard \citep{stiglitz1981credit}. Recent literature extends this framework, introducing the \textit{risk-taking channel} whereby policy rates affect risk premia and banks' willingness to lend \citep{kashyap2023monetary}. Low rates compress risk spreads, potentially building vulnerabilities that surface during tightening episodes, triggering sharp credit contractions and elevated lending standards. Concurrently, a \textit{deposits channel} emerges from imperfect deposit rate pass-through \citep{hannan1997rigidity,drechsler2017deposits}, causing heterogeneous responses based on banks' deposit franchises. \cite{volk2025} documents that, during the ECB's latest tightening cycle, banks increasingly accommodated higher borrower risk by adjusting loan volumes rather than further raising spreads. \citet{de2025banks} show that banks with higher inflation expectations reallocate credit toward highly leveraged firms, highlighting that banks subjective inflation expectations affect the \emph{composition} of lending rather than overall credit supply. In contrast, we focus on the aggregate effects of subjective
inflation uncertainty as a channel of credit rationing, with particular attention to the interest rates banks charge to firms. For this, our paper builds on the classic Monti–Klein model \citep{klein1971theory,monti1972deposit} by incorporating banks' risk aversion, inflation uncertainty, and subjective inflation expectations. These extensions bridge traditional banking theory with contemporary monetary transmission insights, allowing analysis of policy impacts on credit supply, deposit-loan spreads, and bank behavior under uncertainty and expectation biases.

Our empirical choice is coherent with the findings of \citet{lacroix2008assessing}, who shows that loan-level rates are multimodal and that finite Gaussian mixtures parsimoniously fit such distributions by uncovering modes that reflect market segmentation. In particular, for non-financial corporate loans, Lacroix documents strong multimodality and heterogeneity, with high dispersion and multiple modes tied to instrument variety and bank specialization—precisely the type of distinct pricing regimes that motivate our specification with component-specific parameters. Building on this literature, we also relate to \citet{alesina2013women}, who analyze overdraft facilities for Italian microfirms and find persistent heterogeneity in pricing even after extensive controls for borrower and market characteristics. Their study highlights that overdrafts, being short-term, standardized, and central to liquidity management, provide a clean setting to isolate the determinants of credit cost differentials. Following this insight, we perform a dedicated robustness test restricted to \emph{new overdrafts} only. This check allows us to minimize concerns about endogeneity of maturity, collateral, or volume and focus directly on the cost component of bank pricing. By doing so, we can assess whether our main results—linking inflation uncertainty and asymmetry to lending rates—hold in an environment where the contractual structure is virtually homogeneous and the interest rate is most clearly the outcome of banks' own decisions.

    \section{Theoretical Model: Loan Pricing with Subjective Inflation Distribution}

We consider a one-period lending model with a monopolistic, risk-averse bank issuing a standardized nominal loan of unit size \`{a} la \citet{ho1981determinants} extended with default (credit) risk into the spread-setting problem \citep{angbazo1997commercial}. The bank faces \emph{uncertainty} about the probability distribution of an uncertain stochastic variable---here, inflation $\pi$. Let $\mathcal{F} = {F_i : i \in I}$ denote the set of \emph{plausible} probability distributions for $\pi$, where $I$ indexes these candidate distributions, and let $\mu$ be a second-order probability measure defined over $I$. The bank also faces uncertainty about borrower default. The bank maximizes expected von Neumann--Morgenstern utility $U(\cdot)$ of real profits, where $U$ is strictly increasing and concave, reflecting risk aversion. The nominal loan interest rate $R_L$ is set monopolistically. The nominal cost of funds (deposit rate), $R_D(\pi)$, follows a Taylor-type monetary policy rule, $R_D(\pi)=R^*+\rho_\pi(\pi-\pi^*)$, with $\rho_\pi>0$, linking higher expected inflation to higher funding costs. Loans are subject to default risk with zero recovery, wherein default entails no repayment while still incurring funding costs. The default probability $p(R_L,\pi,X)$ is endogenous: it rises with the loan rate ($\partial p/\partial R_L>0$) and is convex ($\partial^2 p/\partial R^2_L > 0$), falls with inflation ($\partial p/\partial \pi < 0$, real long-term debt burden reduction benefit) with decreasing speed ($\partial^2 p/\partial \pi^2 < 0$) \citet{bhamra2023high,de2025banks}, and is elevated in adverse macroeconomic conditions $p(R_L,\pi,1) > p(R_L,\pi,0), \quad X \in \{0,1\}$, where $X=1$ indicates adverse macroeconomic conditions. $X$ is interpreted as an exogenous summary of the broader macroeconomic state (growth, unemployment, financial stress).\footnote{\citet{vansteenberghe2025uncertain} documents that (i) individual-level normalized inflation uncertainty and normalized growth uncertainty are strongly correlated and respond to monetary-policy and financial-stress conditions, and (ii) inflation beliefs exhibit a forward-looking structure consistent with a Neo-Keynesian Phillips curve. This suggests a two-way interaction between inflation SPDs, growth SPDs and macro-financial conditions that could, in a richer model, be endogenized through a joint law of motion for $(\pi,X)$ and the bank’s beliefs about them.} Following \citet{stiglitz1981credit}, higher loan rates worsen borrower quality through adverse selection—safer borrowers exit and riskier applicants enter—and moral hazard, as borrowers shift toward riskier projects. Given these assumptions, the bank chooses the nominal loan rate $R_L$ to maximize the expected utility of its real profit from the loan. The bank's random real profit from the loan depends on repayment outcomes and realized inflation $\pi$. Upon repayment, the bank's nominal net margin $(R_L - R_D(\pi))$ is adjusted for inflation, yielding positive real profit. If the borrower defaults, the bank incurs negative real profit $(- R_D(\pi))$, paying principal and deposit interest without repayment recovery. The bank’s optimization problem can be expressed in two layers. First, conditional on a given candidate distribution $F_i \in \mathcal{F}$ for inflation, expected utility is  

\begin{equation}
\label{eq:Ui_definition}
V_i(R_L)
\;=\;
\int_{-\infty}^{\infty}
\rho(R_L,\pi,X)\,dF_i(\pi),
\end{equation}

where  
\[
\rho(R_L,\pi,X)
=
\bigl(1-p(R_L,\pi,X)\bigr)
U\!\Bigl(\tfrac{R_L-R_D(\pi)}{1+\pi}\Bigr)
+
p(R_L,\pi,X)
U\!\Bigl(-\tfrac{R_D(\pi)}{1+\pi}\Bigr).
\]
The function $\rho(R_L,\pi,X)$ is assumed twice continuously differentiable in its arguments and bounded by an integrable envelope, ensuring the validity of differentiation under the integral sign. Consistent with \citet{boyd2001impact}, $\rho$ is non-increasing and concave in~$\pi$. Second, because the bank is uncertain about which inflation distribution $F_i$ governs outcomes, it forms a second-order expectation over the index set~$I$ of plausible models, weighted by the probability measure~$\mu$:

\begin{equation}
\label{eq:bank_objective}
V(R_L)
\;=\;
\int_{i \in I} V_i(R_L)\,d\mu(i),
\qquad
R_L^{\ast}
\;=\;
\arg\max_{R_L>0} V(R_L).
\end{equation}

The model incorporates two primary channels linking inflation to bank outcomes. First, the \textit{Real Return Channel} captures how inflation directly impacts the real value of nominal payoffs. A higher realized inflation rate $\pi$ diminishes the real net interest margin; specifically, real profits upon repayment, $\frac{R_L - R_D(\pi)}{1+\pi}$, decline as inflation increases, since nominal repayments remain fixed while prices rise. Conversely, in default scenarios, inflation slightly reduces real losses, $\frac{R_D(\pi)}{1+\pi}$, by decreasing the real value of nominal obligations. Second, the \textit{Default Risk Channel} captures inflation's effect on default probabilities. Higher inflation reduces borrowers' real debt burdens, lowering default risk $p(R_L, \pi, X)$, whereas higher loan rates $R_L$ increase debt burdens, raising default likelihood—a Fisherian debt-deflation mechanism \citep{fisher1933debt}. \citet{bhamra2023high} model firms with nominal debt and find asymmetric effects: rising expected inflation decreases credit spreads (default premiums), but falling inflation sharply increases spreads due to nominal rigidities. Similarly, our model emphasizes that banks' loss distributions are skewed, as deflation scenarios generate disproportionately severe defaults compared to symmetric inflationary benefits. We explicitly model this non-linearity by allowing asymmetric subjective distributions $F(\pi)$, reflecting heightened sensitivity to deflationary tail risks, which can be aggravated if their is ambiguity which is modeled via $\phi(\cdot)$. Additionally, adverse macroeconomic conditions (captured by $X=1$) uniformly raise default risk across inflation and interest rate levels. These two channels interact dynamically: higher expected inflation elevates funding costs via the Taylor rule yet simultaneously lowers perceived default risk. The bank evaluates expected real profits by weighting these competing effects according to its subjective inflation distribution $F(\pi)$.

In equilibrium, the bank sets the nominal loan rate \(R_L\) to balance higher interest income against heightened default risk. Formally, if the optimal \(R_L^*\) lies in the interior of the feasible set and all functions are differentiable, then the bank’s first‐order condition equates the marginal expected utility gain from raising \(R_L\) with the associated marginal expected utility loss. To locate \(R_L^*\), we have:
\[
   \frac{\partial V(R_L)}{\partial R_L}
   \;=\;
   0
   \quad\Longrightarrow\quad
   \int_{F \in \mathcal{F}}
     \,\frac{\partial V_F(R_L)}{\partial R_L}
   \, d\mu(F)
   \;=\;
   0.
\]
applying the chain rule  and setting \(\partial V/\partial R_L = 0\) at \(R_L = R_L^*\) yields the first‐order condition:
\begin{equation}\label{eq:FOC}
  \begin{split}
\int_{F \in \mathcal{F}}
\int \Big\{\,\frac{\partial p(R^*_L,\pi,X)}{\partial R_L}\big[U\!\Big(\frac{\,- R_D(\pi)\,}{\,1+\pi\,}\Big) - U\!\Big(\frac{\,R^*_L - R_D(\pi)\,}{\,1+\pi\,}\Big)\big]\\ 
+ \frac{1-p(R^*_L,\pi,X)}{1+\pi} U'\!\Big(\frac{\,R^*_L - R_D(\pi)\,}{\,1+\pi\,}\Big)\Big\} dF(\pi)
\mathrm{d}\mu(F)
   = 0.
  \end{split}
\end{equation}

There are two opposing forces:
\begin{enumerate}
  \item \emph{Positive margin effect}: increasing \(R_L\) raises the per‐loan spread \((R_L - R_D)\) in repayment states.
  \item \emph{Negative default‐risk effect}: since \(\partial p/\partial R_L > 0\), higher rates worsen adverse selection and moral hazard, increasing the probability of non‐repayment.
\end{enumerate}

\paragraph{Interior optimum conditions.}
The loan rate \(R_L\) that solves the bank’s problem in \eqref{eq:bank_objective} will be an interior solution if the default probability exhibits sensible boundary behaviour.  
Specifically, it suffices to assume
\begin{equation}
\label{eq:hazard_limits}
\lim_{R_L\to 0} p(R_L,\pi,X)=0,
\qquad
\lim_{R_L\to \infty} p(R_L,\pi,X)=1,
\end{equation}
so that borrowers always repay at very low rates and always default at very high rates.  
Together with quasi-concavity of the expected utility function \(V(R_L)\) and convexity of the default probability (\(\tfrac{\partial^2 p}{\partial R_L^2}>0\)), these assumptions ensure that the first-order condition \eqref{eq:FOC} delivers a unique interior maximizer \(R_L^\ast\).

\begin{proposition}[Uncertainty-Induced Credit Tightening]
\label{prop:Inflation_Uncertainty_CS}
Let $F$ be a distribution of inflation and $\widetilde F$ a mean-preserving spread (MPS) of $F$. The optimal loan rate under the MPS distribution is strictly higher:
\[
   R_L^*(\widetilde F) \;>\; R_L^*(F).
\]
\end{proposition}

\begin{proposition}[Uncertainty–induced Credit Rationing]
\label{prop:credit_rationing}
Let \(F\) and \(\widetilde{F}\) be subjective beliefs about inflation such that 
\(\widetilde{F}\) is a mean-preserving spread of \(F\). The bank, being risk-averse, 
supplies loans according to \(S_F(R_L)=S(V_F(R_L))\) with \(S'(v)>0\), where 
\(V_F(R_L)\) is the expected utility of real returns. Borrower demand is 
\(D(R_L)\) with \(D'(R_L)<0\).

If under \(F\) the equilibrium rate \(R_L^*\) exhibits credit rationing, 
\(D(R_L^*)>S_F(R_L^*)\), then under \(\widetilde{F}\) the entire supply 
schedule shifts inward, \(S_{\widetilde{F}}(R_L)<S_F(R_L)\) and credit rationing increases for all \(R_L\). 
\end{proposition}

\begin{proposition}[Skewness\mbox{-}Induced Credit Tightening]\label{prop:skew_tightening}
For distributions $F,\widetilde F$ with the same mean, write
$\widetilde F \succeq_{\text{sk}} F$ if
\[
  \mathbb{E}_{\widetilde F}[g(\pi)] \;\le\; \mathbb{E}_F[g(\pi)]
  \quad\text{for all } g:\mathbb{R}\to\mathbb{R} \text{ that are non\mbox{-}increasing and concave,}
\]
with strict inequality for at least one such $g$. We interpret $\widetilde F \succeq_{\text{sk}} F$ as a mean\mbox{-}preserving increase in positive skewness.
Then the bank’s optimal loan rate is (strictly) higher under $\widetilde F$:
\[
   R_L^*(\widetilde F) \;>\; R_L^*(F).
\]
\end{proposition}

\paragraph{Remarks.}
(i) The key economics are: a mean\mbox{-}preserving increase in positive skewness shifts more mass to high\mbox{-}inflation realizations \emph{and} compensating mass to the lower tail; because $\rho$ is decreasing and concave in $\pi$, expected utility at any given $R_L$ falls, while the marginal benefit of raising $R_L$ is larger in higher\mbox{-}$\pi$ states. The FOC therefore tilts toward a higher $R_L$. \\
(ii) An analogous statement holds for a mean\mbox{-}preserving \emph{increase in negative skewness}: it tightens credit at least as much (and typically more) because it directly loads the deflationary tail that banks fear most in this model.

\paragraph{Aggregation to a representative bank.}
Following \citet{vansteenberghe2025insurance}, even in the presence of heterogeneous beliefs across entities, an equivalent representative monopolist can be defined. Let banks be indexed by $b=1,\dots,B$ with loan-share weights $w_{b,t}\ge 0$ in each product--borrower cell at time $t$, $\sum_b w_{b,t}=1$, computed as
\[
w_{b,t} \;=\; \frac{L_{b,t}}{\sum_{j} L_{j,t}}.
\]

Bank $b$ entertains a set of plausible inflation distributions
\[
\mathcal{F}_b \;=\; \{\,F_{b,i} : i \in I_b\,\},
\]
with a second-order measure $\mu_b$ on $I_b$. Conditional on $F_{b,i}$, its expected utility is
\[
V_{b,i}(R_L)
\;=\;
\int_{-\infty}^{\infty}\rho(R_L,\pi,X)\,dF_{b,i}(\pi),
\]
and its second-order objective is
\[
V_b(R_L)
\;=\;
\int_{i\in I_b} V_{b,i}(R_L)\,d\mu_b(i).
\]

Define the aggregate support
\[
\bar{\mathcal{F}}
\;=\;
\bigcup_{b=1}^B \mathcal{F}_b
\;=\;
\{\,F_{b,i} : b=1,\dots,B,\; i\in I_b\,\},
\]
and the mixture measure $\bar\mu$ on $\bar{\mathcal{F}}$ by
\[
\bar\mu(A)
\;=\;
\sum_{b=1}^B
w_{b,t}\;\mu_b\!\big(\{\,i\in I_b : F_{b,i}\in A\,\}\big),
\qquad A\subseteq \bar{\mathcal{F}}.
\]
The representative monopolist then solves
\[
\bar V(R_L)
\;=\;
\int_{F\in \bar{\mathcal{F}}}
\int \rho(R_L,\pi,X)\,dF(\pi)\,d\bar\mu(F)
\;=\;
\sum_{b=1}^B w_{b,t}\,V_b(R_L),
\qquad
R_L^{\text{rep}}
\;=\;
\arg\max_{R_L>0}\,\bar V(R_L).
\]

By linearity, $\bar V(R_L)$ is the loan‐weighted average of banks’ objectives, so $R_L^{\text{rep}}$ summarizes the multi‐bank problem under common primitives $U$, $\rho$, and $R_D(\pi)$; this holds provided banks share identical utility curvature, a common risk–return technology up to fixed effects, and a funding rule differing only by additive bank components.

\subsection{Fixed versus floating-rate loans.}

The formal model is most naturally interpreted as describing a fixed-rate nominal loan. The real return channel operates through the ex post real margin
\(\tfrac{R_L - R_D(\pi)}{1+\pi}\): when inflation turns out higher than expected, the real value of a fixed nominal repayment is eroded and the real net interest margin falls; when inflation is low or negative, the real burden of the fixed payment rises and so does the real margin, at the cost of higher default risk. This mechanism presumes that the contractual loan rate \(R_L\) is predetermined and does not adjust mechanically to inflation realizations. By contrast, for fully floating-rate loans, the contractual rate is periodically reset as a function of policy rates and thus of \(\pi\). In that case, both the numerator (the nominal spread) and the denominator \((1+\pi)\) of the real margin co-move with inflation, and the pure real return channel is largely muted: unexpected inflation is absorbed by contractually higher nominal rates rather than by ex post erosion of a fixed nominal obligation.

The default risk channel must therefore be understood differently for flexible contracts. In the baseline specification, higher inflation lowers default risk, \(\partial p/\partial \pi<0\), because it alleviates the real burden of a fixed nominal debt. For floating-rate loans, however, higher realized inflation is typically associated with higher policy rates and hence higher contractual loan rates at reset dates. In those states, borrowers face an increase in debt service and no longer benefit from the same real-debt relief as under fixed-rate contracts. For sufficiently strong or rapid pass-through, and especially under extreme inflation realizations, this can reverse the sign of \(\partial p/\partial \pi\): higher inflation may increase the probability of default because contractual payments adjust upwards. At the same time, inflation may coincide with higher nominal revenues for firms, which dampens this effect. The overall impact of inflation on default for floating-rate contracts is therefore ambiguous and depends on contract design (caps, floors, reset frequency) and on the joint distribution of inflation and cash flows.

These considerations are particularly relevant for an uncertainty- or ambiguity-averse bank. From the perspective of the lender’s own payoff, floating-rate contracts hedge the real return channel: nominal margins co-move with inflation and reduce exposure to misperceptions about the future price level. At the same time, shifting more interest-rate risk onto borrowers can increase uncertainty about borrower solvency, because default now depends on a more complex interaction between the path of interest-rate resets, firms’ cash flows and macroeconomic conditions. Floating rates can thus reduce ambiguity about the bank’s real return while increasing ambiguity about default risk. As a result, the net effect of higher inflation uncertainty on the relative attractiveness of fixed versus floating contracts is theoretically ambiguous and will in general depend on institutional details and market structure. In what follows, we therefore make no prediction about contract choice and study loan pricing for a generic nominal loan, interpreted as a fixed-rate benchmark, while viewing contract-mix adjustments as an additional margin left for future research.

\subsection{Ambiguity amplifying effect}

To model the ambiguity aversion of banks, we replace the Savage framework \citep{savage1972foundations} with the \emph{smooth ambiguity model} (SAM) of \citet{klibanoff2005smooth}. Preferences are then represented by a two-stage aggregation of utility: the expected utility of outcomes under each $F$ is first computed, and then these expectations are passed through a concave aggregator function~$\phi$, strictly increasing, which reflects \emph{ambiguity aversion}. Now the bank objective becomes
\begin{equation}
\label{eq:bank_objective_SAM}
   \max_{R_L} \; V(R_L)
   \;=\;
   \max_{R_L}\;
   \int_{F\in \mathcal{F}}
     \phi\!\Bigl(V(R_L)\Bigr)
   \, d\mu(F).
\end{equation} where $\phi(\cdot)$ is an increasing, concave ``ambiguity aggregator.'' If $\phi$ is \emph{linear}, the decision-maker is ambiguity-neutral, and $V(R_L)$ coincides with standard SEU---the difference arises when $\phi$ is concave, capturing a preference for caution under model uncertainty. When banks' loss distributions are skewed, the ambiguity aversion modeled via $\phi(\cdot)$ heightens the sensitivity to inflationary tail risks. The FOC becomes
\begin{equation}\label{eq:FOC_SAM}
  \begin{split}
\int_{F \in \mathcal{F}}
\phi'\bigl(V(R_L^*)\bigr)
\int \Big\{\,\frac{\partial p(R^*_L,\pi,X)}{\partial R_L}\big[U\!\Big(\frac{\,- R_D(\pi)\,}{\,1+\pi\,}\Big) - U\!\Big(\frac{\,R^*_L - R_D(\pi)\,}{\,1+\pi\,}\Big)\big]\\ 
+ \frac{1-p(R^*_L,\pi,X)}{1+\pi} U'\!\Big(\frac{\,R^*_L - R_D(\pi)\,}{\,1+\pi\,}\Big)\Big\} dF(\pi)
\mathrm{d}\mu(F)
   = 0.
  \end{split}
\end{equation}
The weight \(\phi'\bigl(U_i(R_L^*)\bigr)\) adjusts each distribution’s contribution: because \(\phi\) is concave, \(\phi'\) is larger when \(U_i(R_L^*)\) is lower, so scenarios with lower expected utility (e.g., deflationary states) receive greater emphasis. Thus, ambiguity aversion induces a “pessimistic” tilt, adding a precautionary premium against uncertainty in inflation and default probabilities. Consequently, the equilibrium loan rate \(R_L^*\) reflects both standard risk aversion and this additional ambiguity‐driven premium.

As \(\phi\) is strictly increasing, the Propositions \ref{prop:Inflation_Uncertainty_CS} and \ref{prop:credit_rationing} still hold and the results are stronger. Meaning that when banks are ambiguity averse, they will raise the bank lending cost and they will ration more their credit supply in times of higher inflation uncertainty.

\paragraph{Orthogonality of Reserve Requirements and Operating Costs}%
\label{par:MK_orthogonality}

Let the real profit in \eqref{eq:Ui_definition} be augmented by  
(i) a non‑remunerated reserve requirement at rate $\theta\in(0,1)$ and  
(ii) a per‑loan operating cost $c\ge 0$, both taken from the
Monti–Klein framework.  
For each inflation realisation $\pi$ define
\[
   \tilde{\rho}(R_L,\pi)
   \;=\;
   \rho(R_L,\pi)
   \;-\;
   \frac{\theta\,R_D(\pi)+c}{1+\pi},
\]
and denote the corresponding expected‑utility objective by
$\tilde V_F(R_L)$ and $\tilde V(R_L)$.

\begin{lemma}[Reserve‑and‑Cost Neutrality]%
\label{lem:orthogonality}
For all $F\in\mathcal{F}$ and all $R_L>0$,
\[
   \frac{\partial\tilde V_F(R_L)}{\partial R_L}
   \;=\;
   \frac{\partial V_F(R_L)}{\partial R_L},  
   \qquad
   \frac{\partial^2\tilde V_F(R_L)}{\partial R_L^2}
   \;=\;
   \frac{\partial^2 V_F(R_L)}{\partial R_L^2}.
\]
Hence the first‑order condition \eqref{eq:FOC} and all ensuing
comparative‑statics results (Propositions
\ref{prop:Inflation_Uncertainty_CS}–\ref{prop:credit_rationing})
remain unchanged after the inclusion of
$\theta$ and $c$.
\end{lemma}

    \section{Empirical Analysis}\label{sec:empirical}

\begin{figure}[h!]
    \centering
    \begin{tabular}{cc}
        \includegraphics[width=0.45\textwidth]{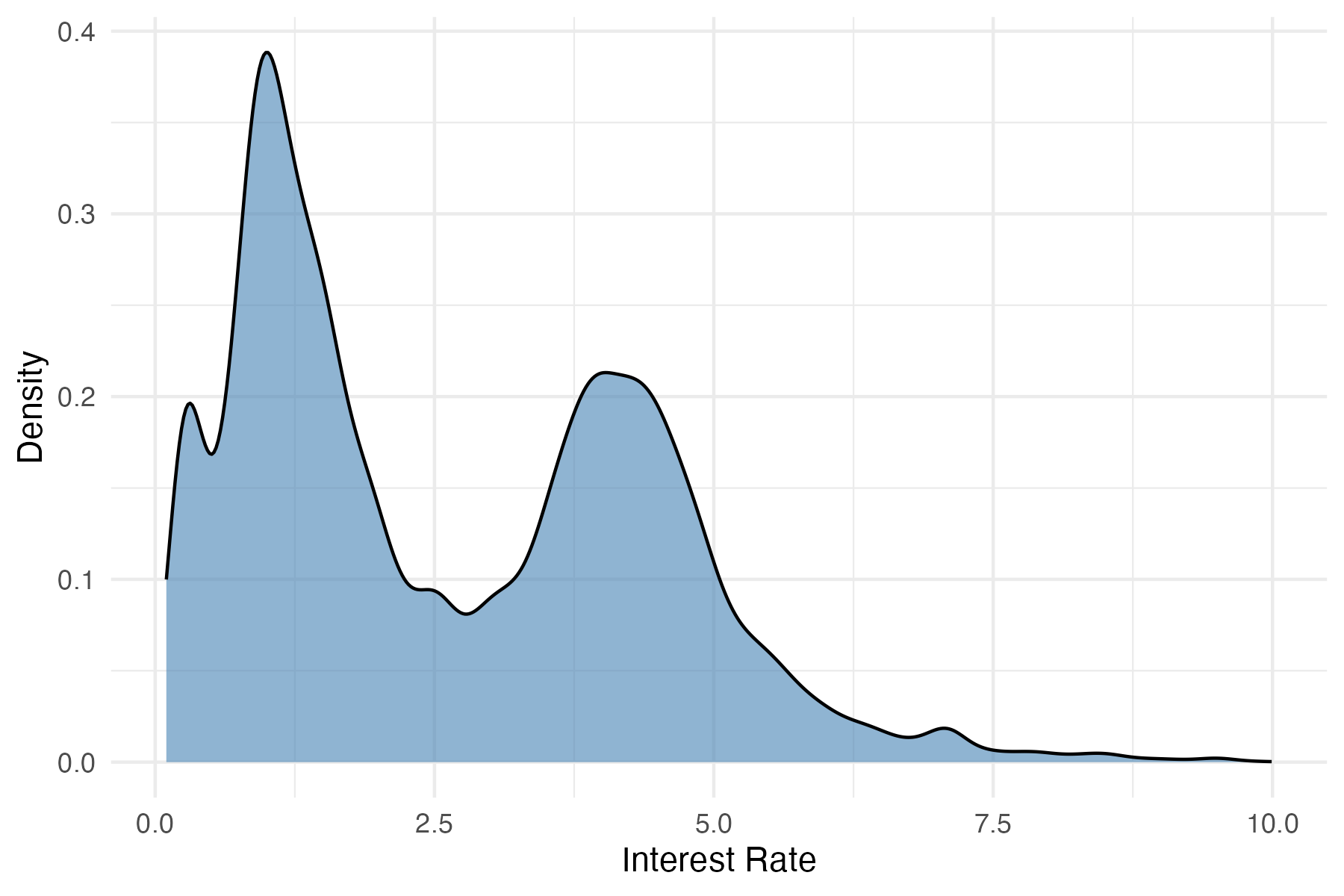} & \includegraphics[width=0.45\textwidth]{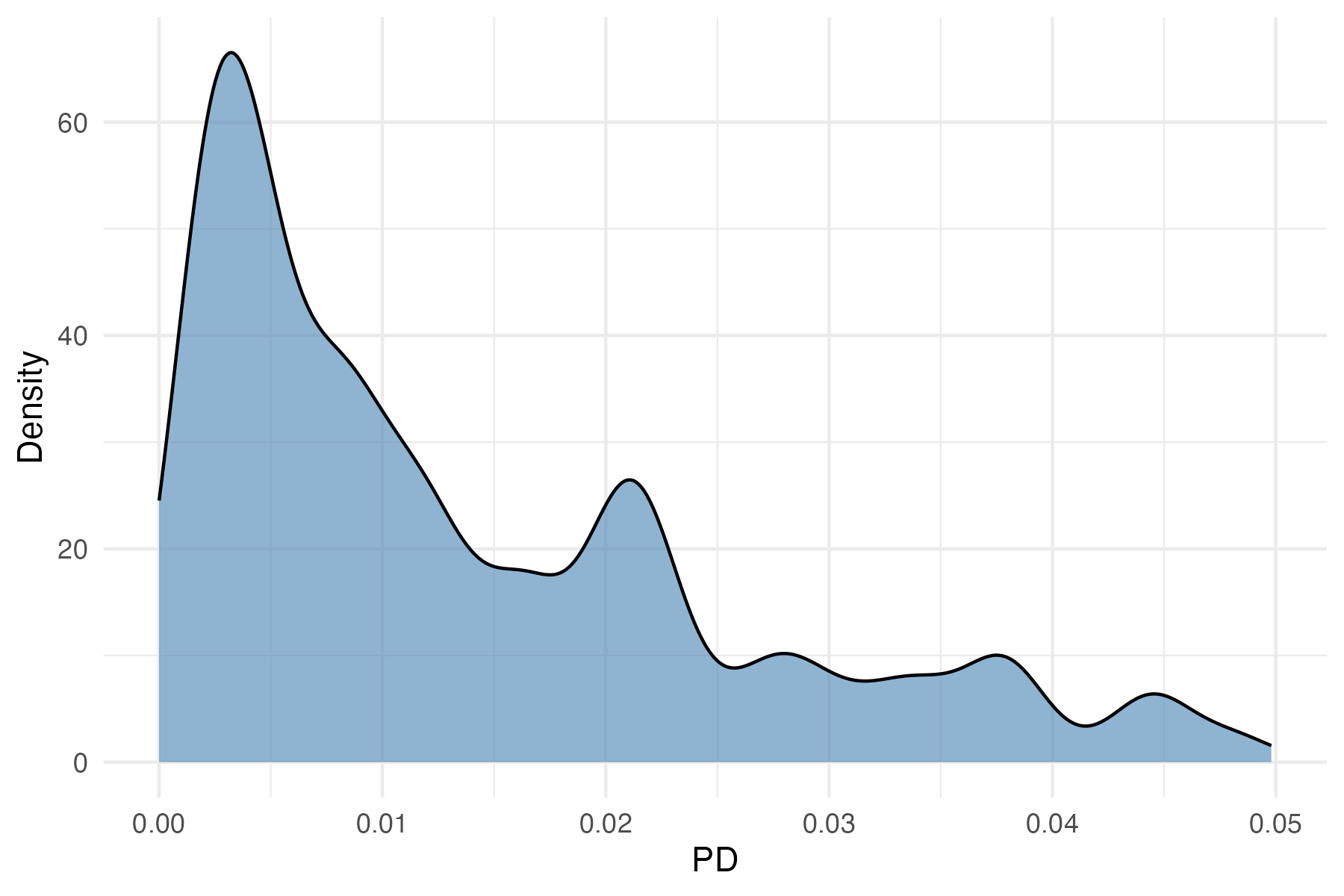} \\
        \includegraphics[width=0.45\textwidth]{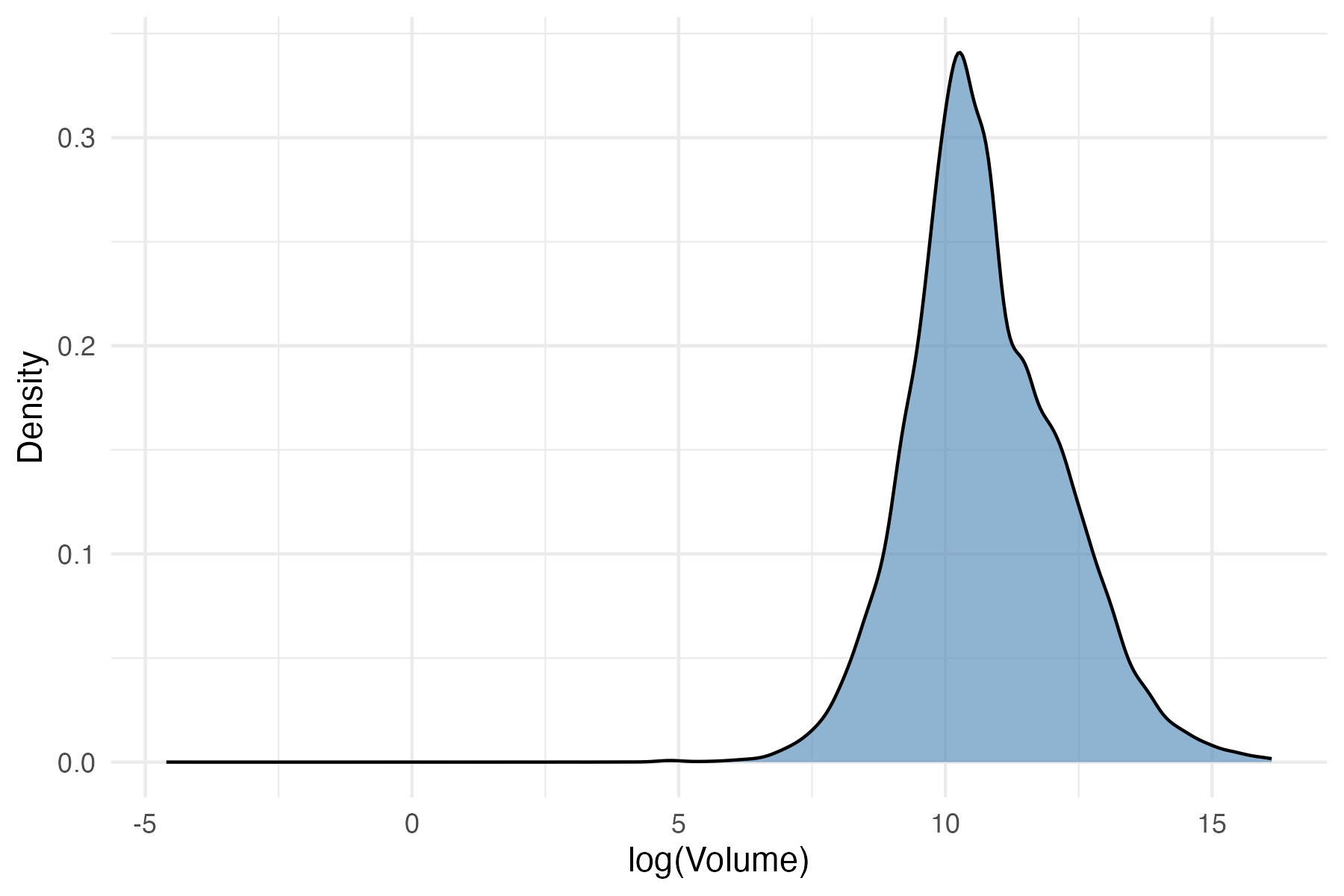} & \includegraphics[width=0.45\textwidth]{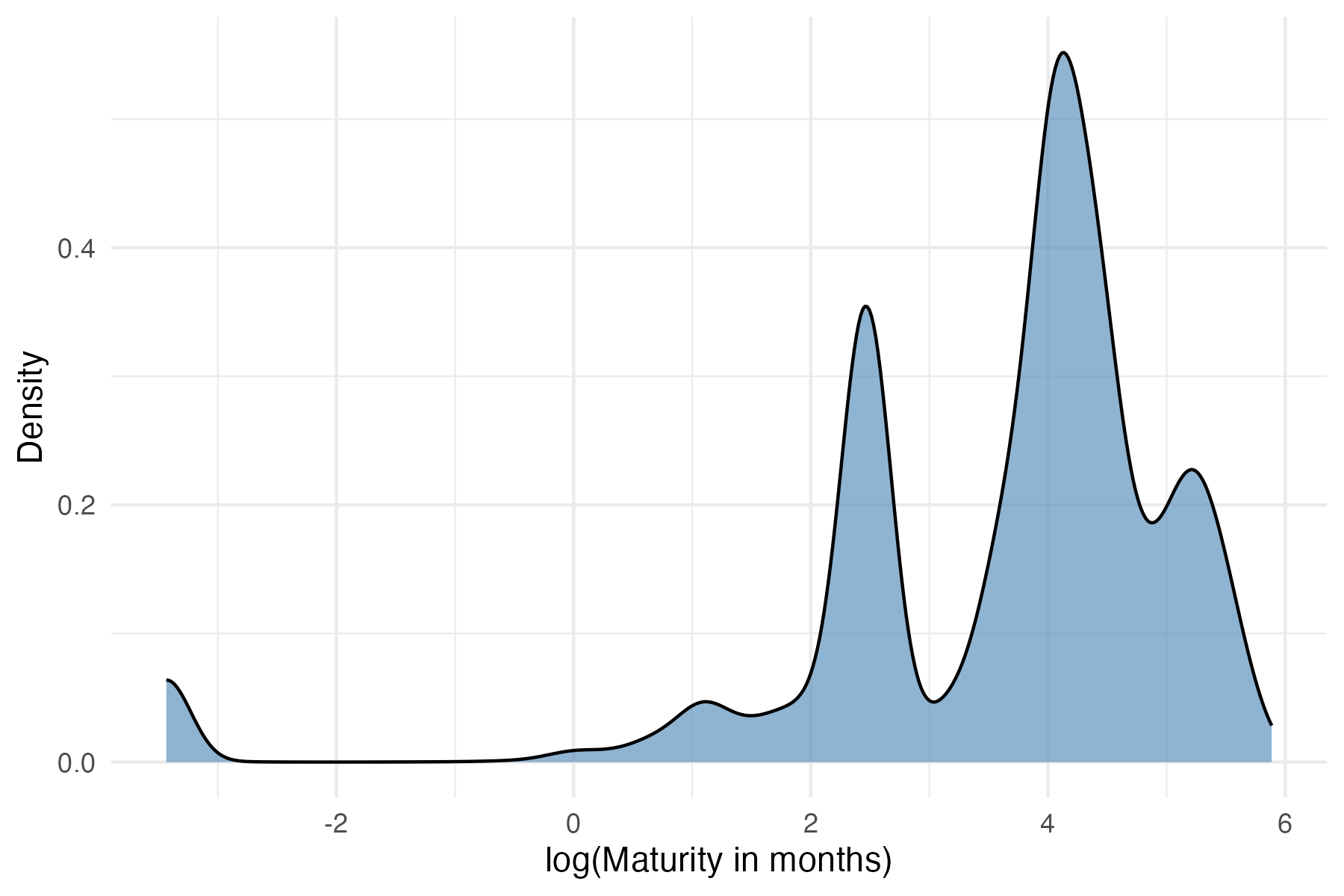} \\
    \end{tabular}
    \caption{Kernel density plots of newly issued bank loans to non-financial corporations in the French market, based on Anacredit data from September 2018 to January 2025. We focus on loans other than overdrafts, convenience credit, extended credit, credit card credit, revolving credit, reverse repurchase agreement, trade receivable and financial lease. Source: Anacredit.}
    \label{fig:loan_KD}
\end{figure}

To identify the impact of inflation uncertainty, forecast disagreement, and asymmetry on bank lending, we employ a likelihood-based density regression framework that controls for the standard determinants of loan interest rates. Following our theoretical model and the existing literature, loan pricing is shaped by maturity, volume, borrower credit risk, macroeconomic conditions and monetary policy, as well as bank-level capital constraints and inflation expectations \citep{graham2008corporate}. All else equal, loans with longer maturities to low-risk firms tend to carry higher interest rates because extending the horizon increases exposure to interest rate, inflation, and solvency risks, and typically requires a maturity or term premium. Consistent with this view, \citet{merton1974pricing} shows that credit yield curves are generally upward-sloping for high-grade issuers, while for riskier firms they may be flat or even downward-sloping. Larger loans may benefit from economies of scale and from the stronger bargaining power of large borrowers, leading to lower spreads. At the same time, very large exposures can create concentration risk for banks, thereby requiring higher returns. In our estimates, we find a negative relationship between loan size and interest rates, significant in at least one mixture component. This suggests that, for most loans, scale economies or borrower quality effects dominate, and larger loans are generally granted at lower rates. A potential concern is that negotiated loan terms, such as volume and maturity, may be jointly determined with the interest rate. Larger loans can reduce rates through bargaining power or scale economies, while lower rates can simultaneously encourage firms to increase borrowing. Such simultaneity risks biasing the estimated effect of loan volume. To address this endogeneity, we include borrower fixed effects, which absorb time-invariant firm-specific attributes such as scale, and conduct robustness checks using firm size measures as additional controls. To account for structural differences across industries, we include sector fixed effects based on the borrower’s primary activity (e.g., manufacturing, retail, services). These controls capture broad sectoral heterogeneity without overfitting and absorb industry-specific risk factors or competitive conditions that systematically influence loan pricing. We measure borrower credit risk using the Probability of Default (PD), which is based on banks’ internal rating models validated under the regulatory framework. These PDs are reported on a through-the-cycle basis, meaning that they do not fully reflect contemporaneous fluctuations in the macroeconomic environment. To account for cyclical variation in credit risk, we therefore augment our specification with a measure of aggregate conditions, proxied by monthly industrial production. In line with risk-based pricing theory, higher PDs should translate into higher loan rates, compensating lenders for expected credit losses and the additional capital charges associated with riskier exposures \citep{wooldridge2010econometric}. We exclude firms with a Probability of Default (PD) above 5\% from our sample, as such “zombies” exhibit atypical lending relationships. As shown by \citet{caballero2008zombie}, banks often evergreen loans to distressed firms or provide subsidized credit, distorting the standard risk–return trade-off. We capture banks’ inflation expectations through three distribution-based measures: the Normalized Inflation Uncertainty (NIU), which isolates individual forecast uncertainty; forecast disagreement across banks; and the Asymmetry Strength Index (ASI), which quantifies skewness in subjective densities. Banks facing higher uncertainty or asymmetry in their inflation outlook are expected to charge a premium on loan rates, both to preserve real returns and to hedge against funding cost risks. We further control for lender fixed effects, $\lambda_{l(i)}$, to account for persistent differences in banks’ pricing strategies and funding costs, such as business model, market power, or borrower composition. Identification thus relies on within-bank deviations in forecast dispersion over time. The significance of NIU, disagreement, and ASI in the presence of bank dummies underscores that it is the time-varying component of inflation expectations, rather than fixed institutional traits, that drives loan pricing outcomes. Figure~\ref{fig:loan_KD} and Table~\ref{tab:desc_anacredit_fixedrate} reports descriptive statistics for the loan sample. The mode of maturity is five years, with an average loan volume of approximately 20,000 euros. The observation period spans both low and high interest rate environments, contributing to the multimodality observed in the loan rate density.

We employ a finite mixture of generalized linear models to capture latent segments or regimes in the data, such as different pricing strategies or borrower groups. The model is estimated via maximum likelihood using an EM algorithm, and the number of components $G$ is selected to balance goodness of fit and parsimony, guided by information criteria. This approach allows us to identify distinct latent structures in the loan data \citep{grun2008flexmix}. Formally, let $r_{i}$ denote the interest rate on loan $i$. Conditional on covariates $x_{i}$, the density of $r_{i}$ is modeled as a mixture of $G$ components:
\begin{equation}\label{eq:mixture}
f(r_{i}\mid x_{i}) \;=\; \sum_{g=1}^{G} \pi_{g}\, f_{g}\!\left(r_{i}\mid x_{i};\,\beta_{g}\right),
\end{equation}
where $\pi_{g}$ is the mixing probability of component $g$ (with $\sum_{g=1}^{G}\pi_{g}=1$), and $f_{g}(r_{i}\mid x_{i};\beta_{g})$ is the likelihood contribution of component $g$, assumed to belong to the GLM family. In our application, we specify $r_{i}$ in each component $g$ as normally distributed with mean $\mu_{ig}=x_{i}'\beta_{g}$ and variance $\sigma_{g}^{2}$. The linear predictor $x_{i}'\beta_{g}$ includes our key covariates—loan characteristics, risk measures, and banks’ forecasts—together with fixed effects. Each component $g$ therefore corresponds to a regression model of $r_{i}$ on $x_{i}$ with its own coefficient vector $\beta_{g}$ and variance $\sigma_{g}^{2}$.

Table~\ref{tab:denregresults} presents the estimation results for the mixture model, with six components labeled as Regime 1 to 6. Across most regimes, the coefficient on \textit{PD} is positive and significant, in line with standard risk-pricing theory. The notable discrepancy arises in regime 1, identified by our model during periods characterized by high GDP growth. This reversal in the sign of the \textit{PD} coefficient is specific to regime 1, where banks seems to search for yield and lend to risky borrowers at affordable rates ceteris paribus. The coefficient on \textit{loan size} is negative in all regimes, underscoring that larger loans are generally associated with slightly lower interest rates, perhaps due to scale economies or stronger bargaining power of large borrowers. The positive and relatively stable effect of the \textit{ECB-DFR} policy rate across all regimes confirms that changes in monetary policy closely pass through into loan pricing. \textit{Maturity} has an ambiguous effect as predicted in the literature. Finally, the positive \textit{GDP} effect across regimes is consistent with macroeconomic conditions influencing borrowing costs: in stronger economic environment, interest rates tend to be higher, possibly reflecting tighter monetary policy or increased demand for credit. The \textit{NIU} (normalized inflation uncertainty index) is also positive in every regime, indicating that higher inflation expectations uncertainty consistently push up nominal lending rates. 

\begin{figure}
    \centering
        \includegraphics[width=0.95\textwidth]{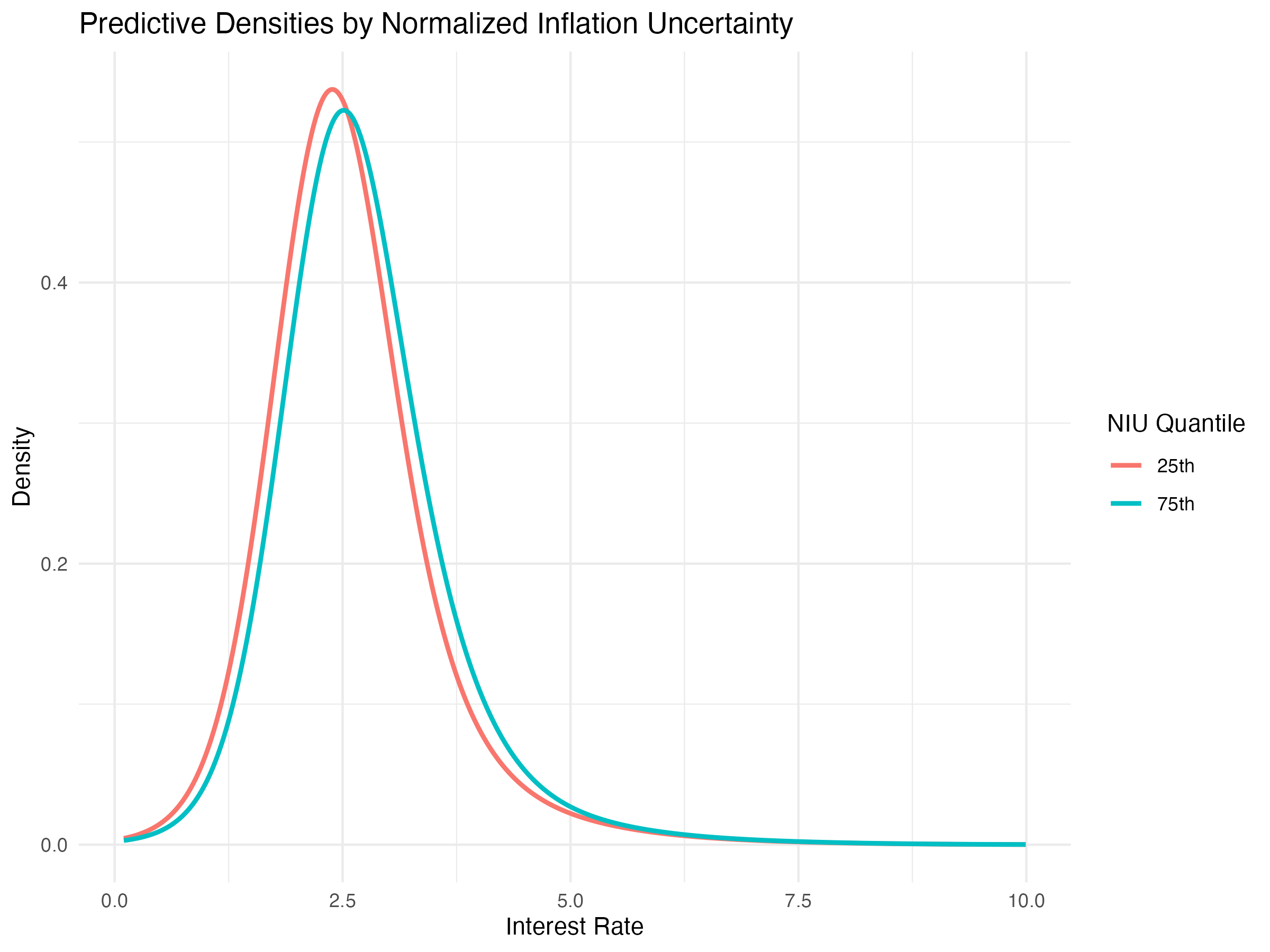} 
    \caption{Predictive densities for newly issued NFC loans, estimated via a 6-component finite mixture model (selected by BIC) holding all controls at their mean and varying the NIU from its 25th to 75th quantile. Source: AnaCredit and author's computation.}
    \label{fig:pred_densityNIU}
\end{figure}

Figure~\ref{fig:pred_densityNIU} and Table~\ref{tab:NIUeffects} summarize the baseline effects of forecast dispersion on loan pricing. Consistent with the theoretical model, higher Normalized Inflation Uncertainty (NIU) shifts the distribution of interest rates charged to non-financial corporations (NFCs) to the right and increases its positive skewness. Quantitatively, the average loan rate rises by 14 basis points, and the effect reaches 16 basis points in the right tail. For scale, this corresponds to nearly half a billion euros in additional annual interest expenses.\footnote{Assuming a monthly flow of 5 billion euros in fixed-rate NFC loans and a maturity mode of five years, the implied stock is roughly 60 billion euros per year times five years, or 300 billion euros. A 14 basis point increase amounts to about \(0.0014 \times 300\) billion \(=\) 0.42 billion euros.} Tables~\ref{tab:DISAGeffects} and \ref{tab:regresultsdisag} report the results for forecast disagreement, and Tables~\ref{tab:ssieffects} and \ref{tab:regresultsSSI} report the results for asymmetry measured by the Asymmetry Strength Index (ASI). Stronger positive skewness in subjective inflation densities increases both the median loan rate and the skewness of the loan-rate distribution, with an effect that is comparable in magnitude to NIU. Since variance and skewness of subjective densities need not be correlated and are in practice not correlated \citep{vansteenberghe2025uncertain}, these findings indicate that the second and third moments carry independent pricing information. By contrast, the effects of forecast disagreement are smaller and less systematic across specifications.

\subsection{Robustness: New Overdrafts as a Natural Laboratory}\label{subsec:overdrafts}

We test the main mechanism on a contract class where maturity, collateral, and volume are not jointly negotiated with price: \emph{new overdrafts}. Overdrafts are short-term, standardized liquidity lines whose pricing is predominantly a bank decision at (re)setting dates, with limited scope for contractual co-determination. This mirrors the identification insight in \citet{alesina2013women}, who exploit overdrafts to study pricing differentials precisely because contractual heterogeneity is minimal. We restrict our robustness analysis to new overdrafts to mitigate selection and simultaneity concerns. Firms that use overdrafts are not randomly selected: NFCs typically resort to these facilities out of short-term liquidity needs. However, overdrafts are standardized contracts with pre-defined conditions agreed upon before any drawdown. A related concern is whether overdraft contracts are genuinely standardized or whether their conditions embed latent renegotiation or risk-sensitive adjustments. Evidence from \citet{martinez2025lines} shows that, for European SMEs, banks adjust overdraft limits, interest rates, and fees in response to firm-specific risk, and overdraft conditions are revised at high frequency. This confirms that overdraft pricing is inherently sensitive to borrower characteristics and often reflects the bank’s ongoing perception of risks. When a firm activates its overdraft, it effectively reveals the terms previously granted by the bank, rather than negotiating them ex post. We therefore interpret observed overdraft rates as direct reflections of banks’ ex ante pricing decisions under standardized contract features. This setup isolates the interest rate spread as the sole relevant margin of adjustment, abstracting from jointly determined loan characteristics such as volume, maturity, or collateral. By focusing exclusively on overdraft conditions, rather than on the amount drawn or the evolution of overdraft users, we exploit a natural setting where contract heterogeneity is minimized and the pricing margin can be interpreted as a clean measure of the bank’s response to uncertainty. Table~\ref{tab:desc_anacredit_overdraft} reports descriptive statistics for the overdrafts sample.

A potential additional concern is that inflation expectations may also shape firms' borrowing decisions, generating selection rather than pure pricing effects. In our baseline, bank-side inflation beliefs enter as treatment variables, and identification relies on within-bank, over-time variation with rich fixed effects. Exploiting \emph{new overdrafts} mitigate selection as contractual co-determination is minimal. Terms are approved \emph{ex ante} and revealed upon first activation; they are not re-negotiated at drawdown. This limits the scope for firms' contemporaneous beliefs about inflation to affect price formation at the moment of use. Because the horizon is short and collateral, maturity, and volume are fixed by design, the bank's perceived inflation risk is the primary margin shaping the spread.

Empirically, equations~\eqref{eq:overdraft_spread} show that NIU remains positive and significant with bank fixed effects and saturated controls (Table~\ref{tab:od_niu}), while ASI and disagreement attenuate once fixed effects are included (Tables~\ref{tab:od_asi}–\ref{tab:od_disag}). These findings align with the baseline mixture estimates and indicate that bank-side inflation uncertainty operates primarily as a \emph{pricing treatment}, not via selection into credit or contract terms.

We recast loan pricing in cost-spread form. Let $r_{i,t}$ be the agreed annualized rate on new overdrafts granted by bank $i$ in month $t$. We estimate parsimonious panel specifications that alternate one uncertainty proxy at a time due to high cross-correlation among NIU, ASI, and disagreement:
\begin{equation}\label{eq:overdraft_spread}
r_{i,t}^{\text{spread}} \;=\; \alpha_i \;+\; \beta_1 \,\mathsf{U}_t \;+\; \delta' X_{i,t} \;+\; \gamma' Y_t \;+\; \varepsilon_{i,t},
\end{equation}
where $\alpha_i$ are bank fixed effects, $\mathsf{U}_t \in \{\text{NIU}_t,\text{ASI}_t,\text{Disagreement}_t\}$ (to avoid multicollinearity), $X_{i,t}$ includes borrower PD and coarse borrower controls (industry, location, firm size), and $Y_t$ collects macro factors (real activity and monetary stance). Standard errors are two-way clustered at the bank and borrower level. Consistent with the structure of the data—newly overdraft facilities with limited repetition at the borrower level—two-way clustering yields virtually identical standard errors to bank-level clustering, indicating that residual dependence is primarily absorbed by bank fixed effects. Identification rests on within-bank, over-time movements in forecast dispersion in a setting where volumes are mechanically tied to short-horizon liquidity management.

The overdraft-only estimates corroborate the baseline:
(i) NIU is positive and statistically significant throughout progressively saturated models with macro controls, bank fixed effects, and borrower-sector-size-department fixed effects (see Table~\ref{tab:od_niu}). Effects are economically close to the full-sample estimates, indicating a robust \emph{cost-setting channel} whereby greater individual inflation uncertainty widens banks’ pricing margins on marginal liquidity to firms.
(ii) ASI loads positively in bivariate models but becomes small and imprecise once bank and rich controls are introduced (Table~\ref{tab:od_asi}), consistent with skewness mattering less for very short-horizon pricing once bank-specific funding and risk conditions are absorbed.
(iii) Disagreement is positive but modest; significance weakens with fixed effects and saturating controls (Table~\ref{tab:od_disag}). This pattern matches the main sample where disagreement conveys less pricing information than NIU.

    \section{Conclusion}

This paper develops and applies distribution-based measures of inflation expectations to study how subjective forecast dispersion shapes the transmission of monetary policy through bank lending. We construct a Normalized Inflation Uncertainty (NIU) indicator that isolates forecasters’ uncertainty from mechanical shifts in expected inflation relative to the two-percent target, complemented by cross-sectional disagreement and an Asymmetry Strength Index (ASI). Embedding these measures in an extended Monti–Klein framework with risk aversion and subjective beliefs yields clear theoretical predictions: higher uncertainty and greater positive skewness in beliefs raise equilibrium lending rates and tighten credit supply, with ambiguity aversion amplifying these effects.

The empirical analysis confirms these predictions. Using loan-level data and a finite-mixture density regression, we document a systematic and positive effect of NIU on the distribution of lending rates to non-financial corporations. Periods of elevated NIU are associated with a rightward shift and greater positive skewness in loan-rate distributions. Moving from the 25th to the 75th percentile of NIU raises the average lending rate by roughly 14 basis points, and up to 16 basis points in the right tail—an effect corresponding to about half a billion euros in additional annual interest costs for French NFCs. ASI displays comparable effects, reinforcing that higher-order moments of subjective distributions carry independent pricing information beyond mean expectations. Disagreement plays a smaller and less consistent role, while placebo tests based on noise proxies fail to reproduce the patterns found for NIU and ASI.

To address potential endogeneity concerns, we conduct a robustness exercise focusing exclusively on new overdrafts—contracts with homogeneous maturities, minimal collateralization, and standardized volumes. This setting effectively removes the main channels of joint determination between price and contract terms, leaving the lending spread as the primary margin of bank decision. The persistence of NIU’s positive effect in this constrained environment supports the interpretation that banks internalize inflation uncertainty as a risk premium in their cost-setting behavior, rather than through selection across maturities, collateral, or loan sizes. Hence, the results capture a causal pricing channel rather than contractual composition effects.

Overall, the findings establish subjective inflation forecast distribution second and third orders as relevant state variables in the monetary transmission mechanism. Inflation uncertainty and asymmetry in beliefs tighten credit conditions. These results imply that the stance and credibility of monetary policy affect not only the level of lending rates but also their dispersion and skewness, through the subjective risk assessments embedded in banks' expectations.

	\newpage
	\bibliographystyle{aea}
	\bibliography{literature}

    \newpage

\section{Proofs}


\subsection*{Proof of existence and characterization of the optimum}

\paragraph{Step 1: Define the marginal integrand and the derivative of \(V\).}
For each inflation outcome \(\pi\) and loan rate \(R_L\), define
\[
g(\pi,R_L)
\;\equiv\;
\frac{\partial \rho(R_L,\pi,X)}{\partial R_L}
=
\frac{\partial p(R_L,\pi,X)}{\partial R_L}\!\Bigl[
U\!\bigl(-\tfrac{R_D(\pi)}{1+\pi}\bigr)
-
U\!\bigl(\tfrac{R_L-R_D(\pi)}{1+\pi}\bigr)
\Bigr]
+
\frac{1-p(R_L,\pi,X)}{1+\pi}\,
U'\!\bigl(\tfrac{R_L-R_D(\pi)}{1+\pi}\bigr).
\]
Under standard envelope conditions (ensuring differentiation under the integral sign),
the derivative of the bank’s objective can be written as
\[
\frac{dV}{dR_L}
=\int_{F\in\mathcal{F}}\int g(\pi,R_L)\,dF(\pi)\,d\mu(F)
=\int G(R_L;F)\,d\mu(F),
\quad
\text{where }G(R_L;F)\equiv \int g(\pi,R_L)\,dF(\pi).
\]

\paragraph{Step 2: Boundary signs of \(g\) and existence of an interior optimum.}
Assume the default probability satisfies
\(\lim_{R_L\to 0}p(R_L,\pi,X)=0\) and \(\lim_{R_L\to\infty}p(R_L,\pi,X)=1\) for all \(\pi\) and \(X\).

- As \(R_L\to 0\): Since \(p\to 0\), the first term in \(g\) vanishes, and we have
  \[
  g(\pi,R_L)\;\to\; \frac{1}{1+\pi}\,U'\!\Bigl(-\tfrac{R_D(\pi)}{1+\pi}\Bigr)\;>\;0,
  \]
  because \(U'\) is positive. Hence \(\tfrac{dV}{dR_L}>0\) when \(R_L\) is near zero.

- As \(R_L\to\infty\): Because \(p\to 1\), the second term in \(g\) drops out, and the first term becomes
  \(\tfrac{\partial p}{\partial R_L}\bigl[U(-R_D/(1+\pi)) - U\bigl((R_L - R_D)/(1+\pi)\bigr)\bigr]\).  
  Since \(U\) is strictly increasing and concave,  
  \(U\bigl((R_L - R_D)/(1+\pi)\bigr)\to \infty\) faster than \(U(-R_D/(1+\pi))\),  
  making the bracket negative.  With \(\tfrac{\partial p}{\partial R_L}>0\), we get \(g(\pi,R_L)<0\) for large \(R_L\).  
  Thus \(\tfrac{dV}{dR_L}<0\) for sufficiently large \(R_L\).

By continuity of \(g\) in \(R_L\) and the change of sign from positive (near zero) to negative (when \(R_L\) is large), the intermediate value theorem implies there is at least one \(R_L^\ast\in (0,\infty)\) such that \(\tfrac{dV}{dR_L}=0\).  If \(V(R_L)\) is quasi-concave, this stationary point is unique.  

\paragraph{Step 3: Concavity in \(R_L\) and verification of a maximum.}
Differentiate \(g\) with respect to \(R_L\):
\[
\frac{\partial g}{\partial R_L}
=
\frac{\partial^2 p}{\partial R_L^2}\Bigl[
U\bigl(-\tfrac{R_D(\pi)}{1+\pi}\bigr)
-
U\bigl(\tfrac{R_L-R_D(\pi)}{1+\pi}\bigr)
\Bigr]
-
\frac{2}{1+\pi}\,\frac{\partial p}{\partial R_L}\,U'\!\bigl(\tfrac{R_L-R_D(\pi)}{1+\pi}\bigr)
+
\frac{1-p(R_L,\pi,X)}{(1+\pi)^2}\,U''\!\bigl(\tfrac{R_L-R_D(\pi)}{1+\pi}\bigr).
\]
Under the usual assumptions, \(U''<0\) (strict concavity of utility) and \(\tfrac{\partial^2 p}{\partial R_L^2}>0\) (convexity of the default probability in \(R_L\)), each term above is non‑positive, with at least one being strictly negative. Consequently,
\[
\frac{\partial g}{\partial R_L}(\pi,R_L) < 0 \quad\text{for all}\;\pi,
\]
and integrating gives
\[
V''(R_L)
= \int_{F\in\mathcal{F}}\int \frac{\partial g}{\partial R_L}(\pi,R_L)\,dF(\pi)\,d\mu(F)
< 0.
\]
Therefore, the stationary point \(R_L^\ast\) is a strict local (and, by quasi‑concavity, global) maximum.  

\paragraph{Step 4: Conclusion.}
Combining the boundary analysis with the second‑order sign, we conclude that there exists a unique interior optimum \(R_L^\ast>0\) solving
\(\int_{F\in\mathcal{F}}G(R_L^\ast;F)\,d\mu(F)=0\), with \(V''(R_L^\ast)<0\).  Hence the bank’s optimal loan rate is strictly inside \((0,\infty)\), and the function \(V(R_L)\) is strictly concave around the optimum.

\subsubsection*{Detailed proof of Proposition~\ref{prop:Inflation_Uncertainty_CS}}

We have the following from the model:
\begin{enumerate}[label=(\roman*)]
  \item \textbf{Risk aversion:} The bank’s utility $U$ is twice continuously differentiable, strictly increasing, and strictly concave.
  \item \textbf{Default risk sensitivity:} The default probability $p(R_L,\pi,X)$ is continuously differentiable in $\pi$ with 
  \[
    \frac{\partial p}{\partial \pi}<0,
    \qquad 
    \frac{\partial^2 p}{\partial \pi^2}\le 0,
  \]
  i.e.\ default probability decreases in inflation and is weakly concave in $\pi$.
  \item \textbf{Funding-cost rule:} The deposit rate $R_D(\pi)$ is affine in inflation, 
  \[
    R_D(\pi) \;=\; R^* + \rho_\pi(\pi-\pi^*), 
    \qquad \rho_\pi>0.
  \]
  (This concavity holds when $U$ is concave, $p$ is decreasing and concave in $\pi$, and $R_D$ is linear.)
  \item \textbf{Unique interior optimum:} $V(R_L)$ is strictly quasi-concave in $R_L$ and satisfies $V''(R_L)<0$ in the neighbourhood of the optimum, ensuring a unique interior solution $R_L^*(F)$.
\end{enumerate}
We will prove the concavity of the marginal-utility integrand: For each fixed $R_L$, the function 
  $\pi \mapsto g(\pi, R_L)$ in the first-order condition is concave in $\pi$.  

\paragraph{1. Setup.}
For a fixed loan rate $R_L$, define
\[
G(R_L;F) \;=\; \int g(\pi,R_L) \, dF(\pi),
\]
where $g(\pi,R_L)$ is the integrand in the first-order condition \eqref{eq:FOC}. By the optimality of $R_L^*(F)$, one has
\[
G\bigl(R_L^*(F);F\bigr) \;=\; 0.
\]
Similarly, $R_L^*(\widetilde F)$ satisfies
\[
G\bigl(R_L^*(\widetilde F);\widetilde F\bigr) \;=\; 0,
\]
where $\widetilde F$ has the same mean as $F$ but is a mean-preserving spread.

\paragraph{2. Concavity of $g(\pi,R_L)$.}
By assumption (iv), for each fixed $R_L$, the function $\pi \mapsto g(\pi,R_L)$ is concave. This concavity is a consequence of the bank’s risk aversion and the default and funding structures: $U$ is strictly concave, $p(R_L,\pi,X)$ is decreasing and concave in $\pi$, and $R_D(\pi)$ is linear in $\pi$. Hence, $g(\pi,R_L)$ inherits concavity from these components.

\paragraph{3. Effect of a mean-preserving spread.}
A mean-preserving spread of $F$ is a distribution $\widetilde F$ that, for some random variable $Z$ with $\mathbb{E}[Z]=0$, satisfies $\widetilde F = \operatorname{Law}(\pi + Z)$ whenever $\pi \sim F$. Because $g(\cdot,R_L)$ is concave in $\pi$, Jensen’s inequality implies
\[
\mathbb{E}_{\widetilde F}[\,g(\pi,R_L)\,] \;=\; \mathbb{E}_F\bigl[g(\pi + Z,R_L)\bigr]
\;\le\; \mathbb{E}_F\bigl[g(\pi,R_L)\bigr],
\]
with strict inequality if $Z$ is non-degenerate and $g$ is strictly concave. In integral form,
\[
G(R_L;\widetilde F)
\;=\;
\int g(\pi,R_L)\, d\widetilde F(\pi)
\;\le\;
\int g(\pi,R_L)\, dF(\pi)
\;=\;G(R_L;F).
\]

\paragraph{4. Comparing first-order conditions at $R_L^*(F)$.}
Evaluate $G(\cdot;F)$ and $G(\cdot;\widetilde F)$ at $R_L = R_L^*(F)$. We have $G(R_L^*(F);F)=0$ by optimality under $F$. From the inequality above,
\[
G\bigl(R_L^*(F);\widetilde F\bigr)
\;\le\;
G\bigl(R_L^*(F);F\bigr)
\;=\;
0.
\]
Because the mean-preserving spread is non-degenerate and $g(\pi,R_L)$ is strictly concave (or at least concave with some strict concave region), we typically get
\[
G\bigl(R_L^*(F);\widetilde F\bigr) \;<\; 0.
\]
That is, at the old optimum $R_L^*(F)$, the derivative of the expected utility with respect to $R_L$ becomes negative under the new (riskier) distribution $\widetilde F$.

\paragraph{5. Monotonicity in $R_L$.}
By assumption (v), $V(R_L)$ is strictly quasi-concave in $R_L$, and $V''(R_L)<0$ near the optimum. This implies $G(R_L;\widetilde F)$ is strictly decreasing in $R_L$ in the relevant region. To restore the first-order condition to zero (i.e., $G(R_L;\widetilde F)=0$), one must increase $R_L$. More formally, since $G(R_L^*(F);\widetilde F)<0$ and $G(R_L;\widetilde F)$ increases with $R_L$ (the derivative of $V$ with respect to $R_L$ increases), there exists a unique $R_L^*(\widetilde F)$ with
\[
R_L^*(\widetilde F) \;>\; R_L^*(F)
\quad\text{and}\quad
G\bigl(R_L^*(\widetilde F);\widetilde F\bigr)\;=\;0.
\]
This completes the proof.

\subsubsection*{Proof of Proposition~\ref{prop:credit_rationing}}

Because \(U\) is strictly increasing and concave in its argument, \(\rho(R_L,\pi,X)\) is a concave function of the random inflation variable \(\pi\) (for each fixed \(R_L\) and \(X\)).  Standard results on second‑order stochastic dominance state that if \(\widetilde{F}\) is a mean‑preserving spread of \(F\), then \(F\) second‑order stochastically dominates \(\widetilde{F}\); hence every risk‑averse decision maker strictly prefers \(F\) to \(\widetilde{F}\).  In particular $V_{\widetilde{F}}(R_L) \le V_F(R_L)$ with strict inequality if \(\widetilde{F}\) is non‑degenerate.  Thus the bank's expected utility at any given interest rate falls when the inflation belief is replaced by its mean‑preserving spread. Combining the monotonicity of \(S\) yields
\[
S_{\widetilde{F}}(R_L)
\;=\;
S\!\Bigl(V_{\widetilde{F}}(R_L)\Bigr)
\;\le\;
S\!\Bigl(V_{F}(R_L)\Bigr)
\;=\;
S_{F}(R_L),
\]
with strict inequality when the mean‑preserving spread is non‑degenerate.

\subsection*{Proof of Proposition~\ref{prop:skew_tightening}}

The marginal effect \(g(\pi,R_L,X)\) is non‑decreasing and convex in \(\pi\).  Intuitively, raising the loan rate \(R_L\) has a smaller adverse effect (or even a larger benefit) when inflation is high: higher \(\pi\) reduces default probabilities and inflation erodes real debt burdens.  Consequently, for a fixed \(R_L\),
\[
\int g(\pi,R_L,X)\,d\widetilde F(\pi)
\;\ge\;
\int g(\pi,R_L,X)\,dF(\pi),
\]
with strict inequality whenever the skewness shift is non‑degenerate and \(g(\cdot,R_L,X)\) is not affine. Let \(R_L^*(F)\) be the unique maximizer of \(V_F(R_L)\).  Then the first‑order condition gives
\[
  \int g(\pi,R_L^*(F),X)\,dF(\pi)
  \;=\;
  \frac{\partial V_F}{\partial R_L}\bigl(R_L^*(F)\bigr)
  \;=\;
  0.
\] but
\[
  \frac{\partial V_{\widetilde F}}{\partial R_L}\bigl(R_L^*(F)\bigr)
  \;>\;
  0.
\] therefore 
\[
  R_L^*(\widetilde F)
  \;>\;
  R_L^*(F),
\]

    \clearpage

\begin{table}[!ht]
\centering
\caption{Descriptive statistics for newly originated fixed-rate loans to NFCs (France)}
\label{tab:desc_anacredit_fixedrate}
\begin{tabular}{lrrrrrr}
\toprule
 & N & Mean & Std.\ Dev. & Min & Median & Max \\
\midrule
Log loan amount & 976{,}351 & 10.7 & 1.4 & -4.6 & 10.6 & 16.1 \\
Log maturity (months) & 976{,}351 & 3.5 & 1.7 & -3.4 & 4.1 & 5.9 \\
Agreed interest rate (\%) & 976{,}351 & 2.6 & 1.8 & 0.1 & 2.1 & 10.0 \\
Debtor PD (\%) & 976{,}351 & 1.4 & 1.2 & 0.0 & 1.0 & 5.0 \\
\bottomrule
\end{tabular}
\vspace{0.25em}
\begin{minipage}{\linewidth}
\footnotesize \textit{Note:} Sample: 2018--09 to 2025--01; average observations per month: 12{,}680. Loan amounts are winsorized at the 99.5th percentile and negative values are set to missing. Interest rates are winsorized at the 0.5th and 99.5th percentiles; observations above 0.30 are rescaled by dividing by 100. Negative maturities are set to missing; maturities above 30 years are flagged and excluded. The sample is restricted to euro-denominated loans to non-financial corporations in France.
\end{minipage}
\end{table}

\begin{table}[!ht]
\centering
\caption{Descriptive statistics for newly drawn overdraft by NFCs (France)}
\label{tab:desc_anacredit_overdraft}
\begin{tabular}{lrrrrrr}
\toprule
 & N & Mean & Std. Dev. & Min & Median & Max \\
\midrule
Log overdraft amount & 55 593 & 7.2 & 3.6 & -0.2 & 6.7 & 15.8 \\
Agreed interest rate (\%) & 55 593 & 9.5 & 8.1 & 0.0 & 14.1 & 19.2 \\
Debtor PD (\%) & 55 593 & 1.2 & 1.2 & 0.0 & 0.7 & 5.0 \\
\bottomrule
\end{tabular}
\vspace{0.25em}
\begin{minipage}{\linewidth}
\footnotesize \textit{Note:} Sample: 2018-09 to 2025-07. Average observations per month: 670. Overdraft amounts are winsorized at the 99.5th percentile and negative values are set to missing. Interest rates are winsorized at the 0.5th and 99.5th percentiles; observations above 0.30 are rescaled by dividing by 100. The sample is restricted to euro-denominated overdrafts to non-financial corporations in France.
\end{minipage}
\end{table}

\begin{table}[!ht]
\caption{Quantile Interest Rates by NIU Scenario (25th vs 75th)}
\label{tab:NIUeffects}
\centering
\begin{tabular}[t]{lrrr}
\toprule
NIU Scenario & q25 & q50 & q75\\
\midrule
25th & 1.985741 & 2.472734 & 3.020092\\
75th & 2.118790 & 2.617200 & 3.186783\\
\bottomrule
\end{tabular}
\end{table}

\begin{table}[!ht]
\centering
\begin{tabular}{llrrrrr}
  \hline
Coefficient & Type & Comp.1 & Comp.2 & Comp.3 & Comp.4 & Comp.5 \\ 
  \hline
  PD & Estimate & -10.82 & 18.03 & 4.93 & 11.34 & 9.45 \\ 
  & Std. Error & 0.86 & 0.35 & 0.09 & 0.13 & 0.12 \\ 
  Loan Size & Estimate & -0.85 & -0.26 & -0.01 & -0.19 & -0.11 \\ 
   & Std. Error & 0.01 & 0.00 & 0.00 & 0.00 & 0.00 \\ 
  ECB-DFR & Estimate & 0.77 & 0.87 & 0.70 & 0.81 & 0.75 \\ 
   & Std. Error & 0.01 & 0.00 & 0.00 & 0.00 & 0.00 \\ 
  NIU & Estimate & 0.13 & 0.11 & 0.11 & 0.23 & 0.26 \\ 
  & Std. Error & 0.02 & 0.01 & 0.00 & 0.00 & 0.00 \\ 
  Maturity & Estimate & 0.11 & -0.00 & -0.02 & 0.03 & -0.01 \\ 
  & Std. Error & 0.01 & 0.00 & 0.00 & 0.00 & 0.00 \\ 
  GDP & Estimate & 6.70 & 4.41 & 0.12 & 1.90 & 1.93 \\ 
   & Std. Error & 0.15 & 0.07 & 0.03 & 0.03 & 0.03 \\ 
   \hline
\end{tabular}
\caption{Estimation Results for the Mixture Model (Estimates and Standard Errors)} 
\label{tab:denregresults}
\end{table}

\begin{table}[!ht]
\caption{Quantile Interest Rates by Disagreement Scenario (25th vs 75th)}
\centering
\begin{tabular}[t]{lrrr}
\toprule
Disagreement Scenario & q25 & q50 & q75\\
\midrule
25th & 2.006134 & 2.484425 & 3.017502\\
75th & 2.057748 & 2.540982 & 3.084771\\
\bottomrule
\end{tabular}
\label{tab:DISAGeffects}
\end{table}

\begin{table}[!ht]
\centering
\begin{tabular}{llrrrrrr}
  \hline
Coefficient & Type & Comp.1 & Comp.2 & Comp.3 & Comp.4 & Comp.5 & Comp.6 \\ 
  \hline
  PD & Estimate & 10.88 & -11.06 & 18.07 & 4.63 & 10.94 & 8.57 \\ 
   & Std. Error & 0.26 & 0.86 & 0.35 & 0.09 & 0.13 & 0.14 \\ 
  Loan volume & Estimate & -0.07 & -0.85 & -0.26 & -0.01 & -0.19 & -0.11 \\ 
   & Std. Error & 0.00 & 0.01 & 0.00 & 0.00 & 0.00 & 0.00 \\ 
  ECB-DFR & Estimate & 0.70 & 0.79 & 0.88 & 0.71 & 0.83 & 0.78 \\ 
   & Std. Error & 0.00 & 0.01 & 0.00 & 0.00 & 0.00 & 0.00 \\ 
  Disagreement & Estimate & 0.02 & 0.17 & 0.20 & 0.09 & 0.23 & 0.36 \\ 
   & Std. Error & 0.01 & 0.02 & 0.01 & 0.00 & 0.00 & 0.00 \\ 
  Maturity & Estimate & -0.19 & 0.11 & -0.00 & -0.02 & 0.03 & 0.02 \\ 
   & Std. Error & 0.00 & 0.01 & 0.00 & 0.00 & 0.00 & 0.00 \\ 
  GDP & Estimate & 0.22 & 6.67 & 4.27 & 0.25 & 2.01 & 1.84 \\ 
   & Std. Error & 0.09 & 0.15 & 0.07 & 0.03 & 0.03 & 0.03 \\ 
   \hline
\end{tabular}
\caption{Estimation Results for the Mixture Model (Estimates and Standard Errors)} 
\label{tab:regresultsdisag}
\end{table}

\begin{table}[htbp]
  \centering
  \caption{Quantile Interest Rates by SSI Scenario}
  \label{tab:ssieffects}
  \begin{tabular}{lrrr}
    \toprule
    ssi            & q25  & q50  & q75  \\
    \midrule
    25th quantile  & 1.69 & 2.17 & 2.70 \\
    75th quantile  & 1.83 & 2.32 & 2.88 \\
    \bottomrule
  \end{tabular}
\end{table}

\begin{table}[ht]
\centering
\begin{tabular}{llrrrrrr}
  \hline
Coefficient & Type & Comp.1 & Comp.2 & Comp.3 & Comp.4 & Comp.5 & Comp.6 \\ 
  \hline
  PD & Estimate & 21.29 & 11.63 & 8.75 & 5.22 & 5.23 & -14.87 \\ 
   & Std. Error& 0.40 & 0.15 & 0.20 & 0.09 & 0.26 & 0.91 \\ 
  Loan Size & Estimate & -0.24 & -0.13 & -0.14 & -0.03 & -0.17 & -0.87 \\ 
   & Std. Error & 0.00 & 0.00 & 0.00 & 0.00 &  & 0.01 \\ 
  ECB-DFR & Estimate & 0.82 & 0.79 & 0.82 & 0.72 & 0.73 & 0.71 \\ 
   & Std. Error & 0.00 & 0.00 & 0.00 & 0.00 & 0.00 & 0.01 \\ 
  SSI & Estimate & 12.60 & 4.83 & 8.22 & 5.90 & 8.56 & 19.71 \\ 
   & Std. Error & 0.49 & 0.17 & 0.23 & 0.10 & 0.29 & 0.95 \\ 
  Maturity & Estimate & 0.01 & -0.13 & 0.07 & 0.00 & 0.08 & 0.13 \\ 
   & Std. Error & 0.00 & 0.00 & 0.00 & 0.00 & 0.00 & 0.01 \\ 
  GDP & Estimate & 4.19 & 2.09 & 1.82 & 1.14 & 1.53 & 6.06 \\ 
   & Std. Error & 0.07 & 0.04 & 0.04 & 0.02 & 0.06 & 0.15 \\ 
   \hline
\end{tabular}
\caption{Estimation Results for the Mixture Model (Estimates and Standard Errors)} 
\label{tab:regresultsSSI}
\end{table}

\begin{table}[!htbp] \centering 
\caption{Impact of Inflation Uncertainty on Overdraft Lending Rates to Non-Financial Corporations} 
  \label{tab:ols_overdraft} 
\begin{tabular}{@{\extracolsep{5pt}}lcccc} 
\\[-1.8ex]\hline 
\hline \\[-1.8ex] 
 & \multicolumn{4}{c}{Dependent variable} \\ 
\cline{2-5} 
\\[-1.8ex] & \multicolumn{4}{c}{Agreed annualized rate (annlsd\_agrd\_rt\_w)} \\ 
 & Uncertainty & + Macro & + Bank FE & + Sector/Size/Dept FE \\ 
\\[-1.8ex] & (1) & (2) & (3) & (4)\\ 
\hline \\[-1.8ex] 
 NIU & 0.013$^{**}$ & 0.011$^{**}$ & 0.002$^{**}$ & 0.002$^{**}$ \\ 
  & (0.006) & (0.005) & (0.001) & (0.001) \\ 
  PD &  & 0.019$^{**}$ & 0.0005$^{***}$ & 0.001$^{***}$ \\ 
  &  & (0.008) & (0.0001) & (0.0001) \\ 
  DFR &  & 0.019$^{***}$ & 0.011$^{***}$ & 0.011$^{***}$ \\ 
  &  & (0.006) & (0.004) & (0.004) \\ 
  GDP &  & $-$0.008$^{**}$ & $-$0.001$^{***}$ & $-$0.001$^{***}$ \\ 
  &  & (0.004) & (0.001) & (0.0005) \\ 
  Constant & 0.082$^{**}$ & 0.082$^{**}$ & 0.010$^{**}$ & $-$0.007$^{*}$ \\ 
  & (0.038) & (0.033) & (0.004) & (0.004) \\ 
 \hline \\[-1.8ex] 
Sector FE & N & N & N & Y \\ 
Size FE & N & N & N & Y \\ 
Dept FE & N & N & N & Y \\ 
Bank FE & N & Y & Y & Y \\ 
Observations & 57,814 & 57,814 & 57,814 & 57,814 \\ 
R$^{2}$ & 0.026 & 0.150 & 0.954 & 0.956 \\ 
Adjusted R$^{2}$ & 0.026 & 0.150 & 0.954 & 0.956 \\ 
\hline 
\hline \\[-1.8ex] 
\textit{Note:}  & \multicolumn{4}{r}{$^{*}$p$<$0.1; $^{**}$p$<$0.05; $^{***}$p$<$0.01} \\ 
\end{tabular} 
\label{tab:od_niu}
\end{table} 

\begin{table}[!htbp] \centering 
\caption{Impact of Inflation Asymmetry on Overdraft Lending Rates to Non-Financial Corporations} 
  \label{tab:od_asi} 
\begin{tabular}{@{\extracolsep{5pt}}lcccc} 
\\[-1.8ex]\hline 
\hline \\[-1.8ex] 
 & \multicolumn{4}{c}{Dependent variable} \\ 
\cline{2-5} 
\\[-1.8ex] & \multicolumn{4}{c}{Agreed annualized rate (annlsd\_agrd\_rt\_w)} \\ 
 & Disagreement & + Macro & + Bank FE & + Sector/Size/Dept FE \\ 
\\[-1.8ex] & (1) & (2) & (3) & (4)\\ 
\hline \\[-1.8ex] 
 ASI & 0.016$^{***}$ & 0.004$^{**}$ & 0.0002 & 0.0002 \\ 
  & (0.005) & (0.002) & (0.0003) & (0.0003) \\ 
  PD &  & 0.018$^{**}$ & 0.0005$^{***}$ & 0.001$^{***}$ \\ 
  &  & (0.008) & (0.0001) & (0.0001) \\ 
  DFR &  & 0.020$^{***}$ & 0.011$^{***}$ & 0.011$^{***}$ \\ 
  &  & (0.007) & (0.004) & (0.004) \\ 
  GDP &  & $-$0.005 & $-$0.001$^{***}$ & $-$0.001$^{***}$ \\ 
  &  & (0.003) & (0.0002) & (0.0002) \\ 
  Constant & 0.082$^{**}$ & 0.082$^{**}$ & 0.008$^{***}$ & $-$0.009$^{***}$ \\ 
  & (0.037) & (0.033) & (0.003) & (0.003) \\ 
 \hline \\[-1.8ex] 
Sector FE & N & N & N & Y \\ 
Size FE & N & N & N & Y \\ 
Dept FE & N & N & N & Y \\ 
Bank FE & N & Y & Y & Y \\ 
Observations & 57,814 & 57,814 & 57,814 & 57,814 \\ 
R$^{2}$ & 0.038 & 0.134 & 0.953 & 0.956 \\ 
Adjusted R$^{2}$ & 0.038 & 0.134 & 0.953 & 0.956 \\ 
\hline 
\hline \\[-1.8ex] 
\textit{Note:}  & \multicolumn{4}{r}{$^{*}$p$<$0.1; $^{**}$p$<$0.05; $^{***}$p$<$0.01} \\ 
\end{tabular} 
\end{table}

\begin{table}[!htbp] \centering 
\caption{Impact of Inflation Disagreement on Overdraft Lending Rates to Non-Financial Corporations} 
  \label{tab:od_disag} 
\begin{tabular}{@{\extracolsep{5pt}}lcccc} 
\\[-1.8ex]\hline 
\hline \\[-1.8ex] 
 & \multicolumn{4}{c}{Dependent variable} \\ 
\cline{2-5} 
\\[-1.8ex] & \multicolumn{4}{c}{Agreed annualized rate (annlsd\_agrd\_rt\_w)} \\ 
 & Disagreement & + Macro & + Bank FE & + Sector/Size/Dept FE \\ 
\\[-1.8ex] & (1) & (2) & (3) & (4)\\ 
\hline \\[-1.8ex] 
 Disagreement & 0.007$^{*}$ & 0.011$^{**}$ & 0.002$^{*}$ & 0.001$^{*}$ \\ 
  & (0.004) & (0.004) & (0.001) & (0.001) \\ 
  PD &  & 0.019$^{**}$ & 0.0005$^{***}$ & 0.001$^{***}$ \\ 
  &  & (0.008) & (0.0001) & (0.0001) \\ 
  DFR &  & 0.023$^{***}$ & 0.012$^{***}$ & 0.011$^{***}$ \\ 
  &  & (0.007) & (0.004) & (0.004) \\ 
  GDP &  & $-$0.008$^{*}$ & $-$0.001$^{***}$ & $-$0.001$^{***}$ \\ 
  &  & (0.004) & (0.0004) & (0.0003) \\ 
  Constant & 0.082$^{**}$ & 0.082$^{**}$ & 0.009$^{**}$ & $-$0.008$^{**}$ \\ 
  & (0.038) & (0.033) & (0.004) & (0.004) \\ 
 \hline \\[-1.8ex] 
Sector FE & N & N & N & Y \\ 
Size FE & N & N & N & Y \\ 
Dept FE & N & N & N & Y \\ 
Bank FE & N & Y & Y & Y \\ 
Observations & 57,814 & 57,814 & 57,814 & 57,814 \\ 
R$^{2}$ & 0.007 & 0.150 & 0.954 & 0.956 \\ 
Adjusted R$^{2}$ & 0.007 & 0.150 & 0.954 & 0.956 \\ 
\hline 
\hline \\[-1.8ex] 
\textit{Note:}  & \multicolumn{4}{r}{$^{*}$p$<$0.1; $^{**}$p$<$0.05; $^{***}$p$<$0.01} \\ 
\end{tabular} 
\end{table}

    \clearpage

    \FloatBarrier
\pdfbookmark[0]{Appendix}{Appendix}
\appendix
\setcounter{footnote}{0}
\setcounter{table}{0}
\setcounter{figure}{0}
\setcounter{equation}{0}
\setcounter{page}{0}
\renewcommand{\thepage}{\arabic{page}}
\renewcommand{\thesection}{\Alph{section}}
\renewcommand{\thesubsection}{\thesection.\arabic{subsection}}
\renewcommand\thefigure{\thesection.\arabic{figure}}
\renewcommand\thetable{\thesection.\arabic{table}}
\renewcommand\theequation{\thesection.\arabic{equation}}
\numberwithin{table}{section}
\numberwithin{figure}{section}

\makeatletter
\renewcommand\@makefnmark{%
	\hbox{\hspace{-4pt}\@textsuperscript{\normalfont\@thefnmark}}}
\makeatother

\thispagestyle{empty}

\section{Appendix}

\subsection{Robustness check: placebo tests}\label{app:robustness}

As a robustness check, we construct a placebo time series that mimics the statistical properties of our normalized inflation uncertainty measure but carries no economic content. Specifically, we generate a Gaussian white noise process where the mean and variance are set equal to the sample mean and variance of the normalized inflation uncertainty. This placebo series thus preserves the unconditional first and second moments of the true indicator, while breaking any potential relationship with interest rates. 

Replacing the normalized inflation uncertainty with this placebo variable in our regressions, we confirm that the estimated effects disappear. The quantile analysis in Table~\ref{tab:placebo_quantiles} shows no systematic differences, while the mixture regression estimates reported in Table~\ref{tab:regresults} indicate that the placebo variable is statistically insignificant across all components. This reassures us that the main findings are not driven by spurious correlations or by the particular structure of our estimation strategy. 

\begin{table}[ht]
\caption{Quantile Interest Rates by Placebo Scenario (25th vs 75th)}
\label{tab:placebo_quantiles}
\centering
\begin{tabular}[t]{lrrr}
\toprule
Placebo Scenario & q25 & q50 & q75\\
\midrule
25th & 2.063296 & 2.562205 & 3.130395\\
75th & 2.063008 & 2.561927 & 3.130058\\
\bottomrule
\end{tabular}
\end{table}

\begin{table}[ht]
\centering
\begin{tabular}{llrrrrr}
  \hline
Coefficient & Type & Comp.1 & Comp.2 & Comp.3 & Comp.4 & Comp.5 \\ 
  \hline
PD & Estimate & -10.51 & 17.91 & 4.32 & 10.46 & 8.95 \\ 
   & Std. Error & 0.86 & 0.35 & 0.09 & 0.13 & 0.12 \\ 
Volume & Estimate & -0.85 & -0.26 & -0.01 & -0.19 & -0.10 \\ 
   & Std. Error & 0.01 & 0.00 & 0.00 & 0.00 &  \\ 
ECB-DFR & Estimate & 0.78 & 0.87 & 0.71 & 0.81 & 0.75 \\ 
   & Std. Error & 0.01 & 0.00 & 0.00 & 0.00 & 0.00 \\ 
Placebo & Estimate & 0.01 & 0.00 & -0.00 & -0.00 & 0.00 \\ 
   & Std. Error & 0.02 & 0.01 & 0.00 & 0.00 & 0.00 \\ 
Maturity & Estimate & 0.10 & 0.00 & -0.02 & 0.03 & -0.05 \\ 
   & Std. Error & 0.01 & 0.00 & 0.00 & 0.00 & 0.00 \\ 
GDP & Estimate & 7.01 & 4.66 & 0.78 & 2.65 & 2.97 \\ 
   & Std. Error & 0.15 & 0.07 & 0.03 & 0.03 & 0.03 \\ 
   \hline
\end{tabular}
\caption{Estimation Results for the Mixture Model (Estimates and Standard Errors)} 
\label{tab:regresults}
\end{table}

\begin{center}
    \includegraphics[scale=.5]{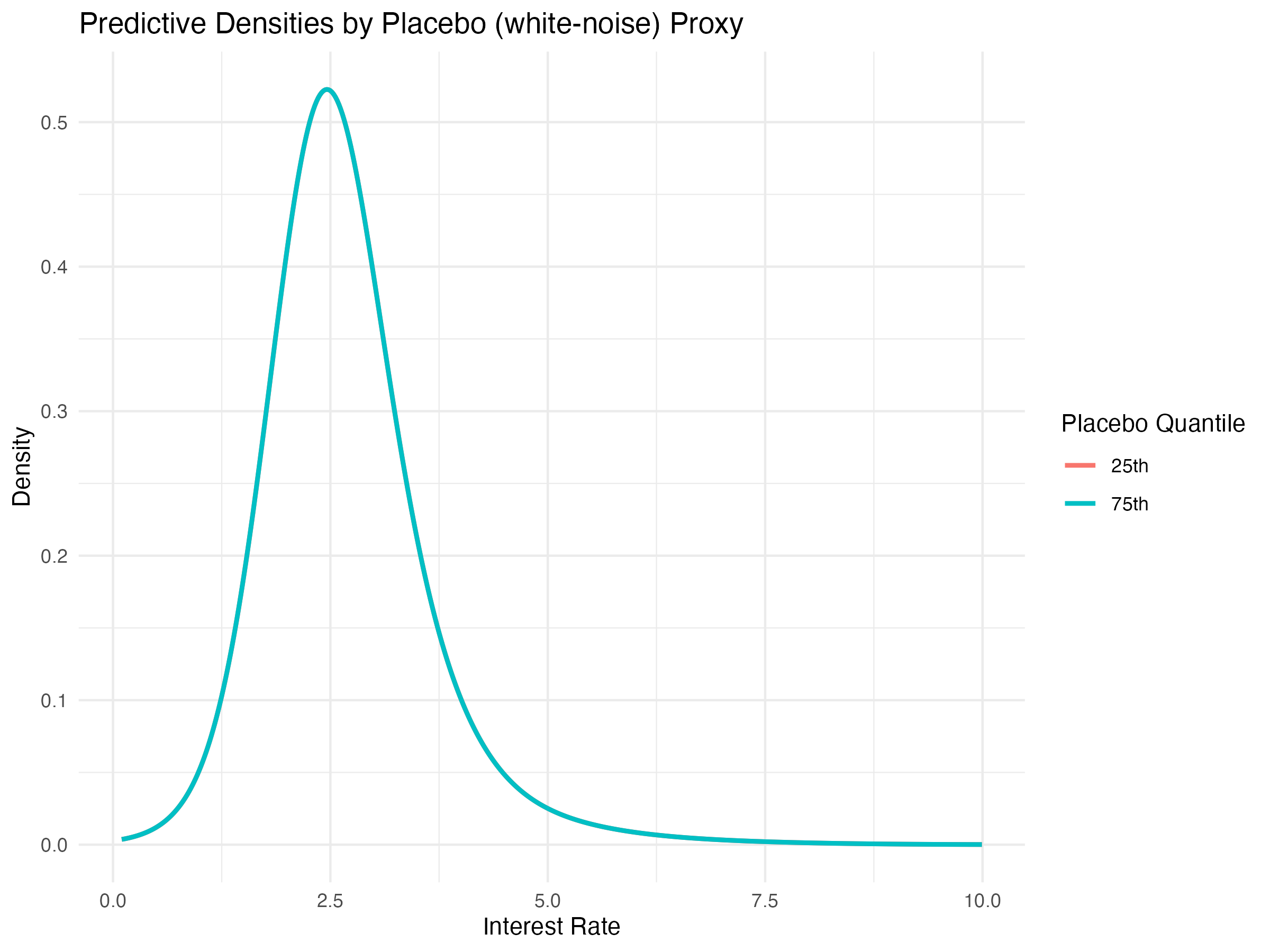}
\end{center}

\end{document}